\documentclass[12pt,preprint]{aastex}

\shorttitle{NIR SPECTROSCOPY OF LIRGS}
\shortauthors{LEE ET AL.}

\begin{document}

\title{{\it AKARI} NEAR-INFRARED SPECTROSCOPY OF LUMINOUS INFRARED GALAXIES}

\author{Jong Chul Lee$^{1,2}$, Ho Seong Hwang$^{3,4}$, Myung Gyoon Lee$^{1}$, 
        Minjin Kim$^{2,5}$, and Joon Hyeop Lee$^{2}$}

\affil{$^{1}$ Astronomy Program, Department of Physics and Astronomy, 
              Seoul National University, Seoul 151-742, Republic of Korea;
              mglee@astro.snu.ac.kr\\
       $^{2}$ Korea Astronomy and Space Science Institute, Daejeon 305-348, 
              Republic of Korea; jclee@kasi.re.kr, mkim@kasi.re.kr, jhl@kasi.re.kr\\
       $^{3}$ CEA Saclay/Service d'Astrophysique, F-91191 Gif-sur-Yvette, France\\
       $^{4}$ Smithsonian Astrophysical Observatory, 60 Garden Street, Cambridge, 
              MA 02138, USA; hhwang@cfa.harvard.edu\\
       $^{5}$ National Radio Astronomy Observatory, 520 Edgemont Road, Charlottesville, 
              VA, USA\\}

\begin{abstract}
We present the {\it AKARI} near-infrared (NIR; 2.5--5 $\mu$m) spectroscopic study 
  of 36 (ultra)luminous infrared galaxies [(U)LIRGs] at $z=0.01-0.4$.
We measure the NIR spectral features including
  the strengths of 3.3 $\mu$m polycyclic aromatic hydrocarbon (PAH) emission and  
  hydrogen recombination lines (Br$\alpha$ and Br$\beta$), 
  optical depths at 3.1 and 3.4 $\mu$m, and NIR continuum slope.
These spectral features are used to identify optically elusive, buried AGN.
We find that half of the (U)LIRGs optically classified as non-Seyferts
  show AGN signatures in their NIR spectra.
Using a combined sample of (U)LIRGs with NIR spectra in the literature,
  we measure the contribution of buried AGN to the infrared luminosity
  from the SED-fitting to the {\it IRAS} photometry.
The contribution of these buried AGN to the infrared luminosity   
  is 5--10\%, smaller than the typical AGN contribution
  of (U)LIRGs including Seyfert galaxies (10--40\%).
We show that NIR continuum slopes correlate well with 
  {\it WISE} [3.4]--[4.6] colors, which would be useful for identifying 
  a large number of buried AGN using the {\it WISE} data.
\end{abstract}



\section{INTRODUCTION}
Since the {\it Infrared Astronomical Satellite} \citep[{\it IRAS};][]{neu84} 
  first opened the all-sky view of far-infrared universe, 
  a large number of luminous infrared galaxies 
  (LIRGs; 10$^{11} \leqq L_{\rm IR}$(8--1000 $\mu$m) $<~$10$^{12}~L_{\odot}$)
  and ultraluminous infrared galaxies (ULIRGs; $L_{\rm IR} \geqq$ 10$^{12}~L_{\odot}$)
  have been identified and studied extensively
  (see \citealt{san96,lon06,soi08} for review).

In the local Universe,
  many of them are interacting systems between gas-rich disk galaxies 
  [e.g., \citealt{kim02,vei02,wan06,kav09,hwa10a}; 
  but see \citealt{elb07,lot08,ide09,kar10,kar11} for high-$z$ (U)LIRGs].
They may evolve into quasars and then into intermediate-mass elliptical galaxies 
  \citep[e.g.,][]{san88a,kor92,gen01,tac02,das06,vei09b,rot10,haa11}.
Their enormous infrared luminosity comes 
  from cool dust primarily heated by young stars [i.e., star formation (SF)] and
  hot dust heated by supermassive black holes (SMBHs) rapidly accreting matter 
  [i.e., active galactic nuclei (AGN)].
Their contribution to the infrared luminosity density
  increases with redshift \citep[e.g.,][]{lef05,mag09,got11}.
Therefore, the study of (U)LIRGs allows us to better understand 
  galaxy-galaxy interactions, starburst-AGN connection, and 
  cosmic star formation history.
  
To identify the energy sources of galaxies (i.e., SF vs. AGN),
 the optical line ratios sensitive to the photoionization source have been used 
 \citep[e.g.,][]{bal81,vei87,kew06}.
Based on this method,
  the optical spectral types for a large number of (U)LIRGs 
  are also determined
  \citep[e.g.,][]{vei95,vei99a,kew01,got05,cao06,hou09,lee11}.
However, the optical spectral classification can be uncertain, 
  in particular for (U)LIRGs,
  due to the difficulty of detecting dust-enshrouded AGN.
In this case, near-infrared (NIR) spectroscopy, less affected by 
  dust extinction, is a more efficient tool.
  
There are several ground-based $K$-band (1.9--2.4 $\mu$m) spectroscopic surveys 
  searching for obscured AGN signatures in (U)LIRGs: 
  the presence of broad Pa$\alpha$ emission centered at (rest-frame) 1.875 $\mu$m
  or of high-excitation coronal line [\ion{Si}{6}] at 1.963 $\mu$m
  \citep[e.g.,][]{vei97,vei99b,mur99,mur01}.
The ground-based $L$-band (2.8--4.2 $\mu$m) spectroscopy was also used to
  constrain their energy source 
  \citep[e.g.,][]{ima00,ima03,ima06,ima11,ris06,ris10,san08}.
The SF-dominated (U)LIRGs show a strong emission line 
  at 3.29 $\mu$m attributed to polycyclic aromatic hydrocarbons (PAHs),
  whereas AGN-dominated (U)LIRGs show a relatively PAH-free continuum 
  attributed to larger-sized hot dust grains \citep[e.g.,][]{moo86,ima00}.
The strong absorption features at 3.05 and 3.4 $\mu$m by H$_{2}$O ice-covered dust grains 
  and bare carbonaceous dust, respectively, are found in (U)LIRGs with buried AGN. 
However, these absorptions are weak or absent in normal SF (U)LIRGs 
  where energy sources and dust are often spatially well mixed \citep[e.g.,][]{ima00,ima03}.
For a similar reason, 
  NIR continua of AGN-dominated (U)LIRGs can be much redder than 
  those of SF-dominated (U)LIRGs \citep[e.g.,][]{ris06}.

Thanks to the wide wavelength coverage (2.5--5 $\mu$m) and 
  high sensitivity of the Infrared Camera (IRC) 
  on-board the {\it AKARI} space telescope \citep[][]{mur07,ona07}, 
  the $L$-band diagnostic could be applied to 
  more distant and faint galaxies \citep[e.g.,][]{ima08,ima10a}.
The AGN detection rate from $L$-band spectroscopy appears to be larger than  
  that from $K$-band spectroscopy (roughly 70\% vs. 20\% in ULIRGs)  
  because the $K$-band diagnostic detects only obvious AGN.
These NIR spectroscopic studies of (U)LIRGs suggest that
  there are many optically elusive buried AGN and that 
  the buried AGN fraction increases with increasing infrared luminosity.    

In this study, we analyze the {\it AKARI} NIR spectra of 
  36 (U)LIRGs\footnote{Although 
  5 out of 36 galaxies do not satisfy the definition of (U)LIRGs 
  (i.e., $L_{\rm IR} <$ 10$^{11}~L_{\odot}$),
  they are simply referred as (U)LIRGs in this study.
  We do not include them when we compare our results 
  with those in previous studies.}
  mainly from the cross-correlation between the {\it IRAS} and 
  Sloan Digital Sky Survey \citep[SDSS;][]{yor00}.
Combining our new NIR spectroscopic data with those in \citet{ima08,ima10a},
  we investigate the NIR properties of a large sample of (U)LIRGs.
The structure of this paper is as follows.
The target selection is explained in Section 2.
Observations and data reduction are described in Section 3,
and the method of NIR spectral analysis is given in Section 4.
Our findings are discussed and summarized in Sections 5 and 6, respectively.
Throughout this paper, we adopt flat $\Lambda$CDM cosmological parameters:
  $H_{0}$ = 75 km s$^{-1}$ Mpc$^{-1}$, $\Omega_{M}$ = 0.3, and $\Omega_{\Lambda}$ = 0.7.

\section{TARGETS}

We obtained the {\it AKARI} NIR spectra of (U)LIRGs in three {\it AKARI} open-time programs:
  the nature of new ULIRGs at intermediate redshift (NULIZ), 
  NIR spectroscopy of composite and LINER LIRGs (CLNSL), and 
  NIR spectroscopy of star-forming infrared galaxies (NISIG).
The main targets for the NULIZ program were selected from a catalog of $\sim$320 ULIRGs 
  in \citet{hwa07}, which was constructed by cross-correlating 
  the {\it IRAS} faint source catalog \citep{fsc92} 
  with the galaxy redshift survey catalogs, including the
  SDSS Data Release 4 \citep[SDSS DR4;][]{ade06}, 
  2dF Galaxy Redshift Survey \citep[2dFGRS;][]{col01}, and 
  6dF Galaxy Survey \citep[6dFGS;][]{jon04,jon05}.
These ULIRGs are detected up to $z \sim$0.4.
Because most ULIRGs observed with $L$-band spectrograph are at $z < 0.2$, 
  we focus on 20 newly discovered ULIRGs at 0.2 $< z <$ 0.4 to extend the NIR spectroscopy 
  to intermediate-$z$ ULIRGs.
The upper redshift limit ensures that the 3.3 $\mu$m PAH emission falls  
  within the AKARI spectral coverage.
Additionally, four nearby ULIRGs were selected, as a control sample,
  from \citet{soi89}, \citet{lee94}, \citet{sta00}, and \citet{san03}.   

For the CLNSL and NISIG programs, 
  we first identified $\sim$14,000 infrared galaxies by the cross-correlation 
  of infrared sources in the {\it IRAS} faint source catalog with 
  the spectroscopic sample of galaxies in the SDSS DR7 
  (\citealt{aba09}; see \citealt{hwa10a} for more details).
We selected $\sim$13,000 local infrared galaxies 
  at 0.01 $< z <$ 0.2.
We then determined their optical spectral types 
  [star-forming, (SF-AGN) composite, 
  low ionization nuclear emission-line region (LINER), and Seyfert 2]
  using their emission line fluxes
  based on the criteria of \citet{kew06}
  (see \citealt{lee11} for more details).
The line flux measurements were drawn from
  the Max-Planck-Institute for Astrophysics/Johns Hopkins University
  value-added galaxy 
  catalog\footnote{http://www.mpa-garching.mpg.de/SDSS/DR7/} 
  \citep[MPA/JHU VAGCs;][]{kau03a,tre04,bri04}.
We finally selected 60 $K_s$-bright\footnote{$K_s$-band (2.17 $\mu$m) 20 mag arcsec$^{-2}$ 
  isophotal ellipse aperture magnitude drawn from the 2 micron all-sky survey (2MASS) 
  $<$ 14.5 mag} non-Seyfert (i.e., SF, composite, and LINER) galaxies,   
  covering a wide range of infrared luminosities 
  ($L_{\rm IR}$ = $10^{10}$--10$^{13}~L_{\odot}$).
Higher priority was assigned
  to $K_s$-bright galaxies in order to obtain high signal-to-noise ratio (S/N) {\it AKARI} spectra
  and to non-Seyfert galaxies in order to find ``buried" AGN.
Most targets are at high ecliptic latitudes 
  due to the limited sky visibility of {\it AKARI} satellite 
  (Sun-synchronous orbit; \citealt{mur07}),
  but this does not introduce any bias to the results.
  
\begin{figure}
\begin{center}
\includegraphics[angle=0,scale=1.]{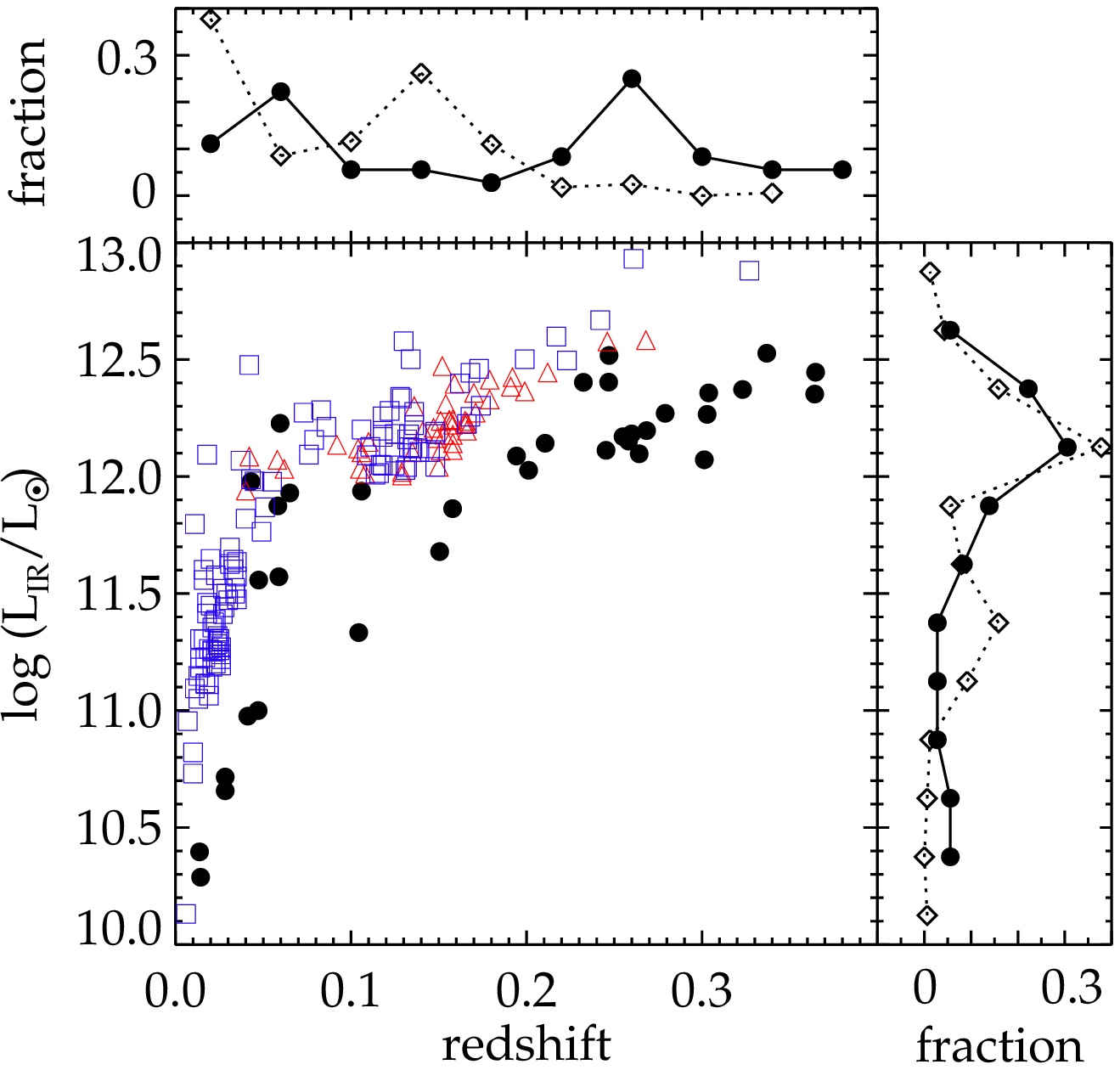}
\caption{
Infrared luminosity vs. redshift distribution of (U)LIRGs observed with {\it AKARI} IRC.
The (U)LIRGs in this study, \citet{ima08}, and \citet{ima10a} are represented 
by black filled circles, red open triangles, and blue open squares, respectively.
In the upper and right panels, the redshift and luminosity distributions of (U)LIRGs are shown, respectively 
(our sample: solid line with filled circles; Imanishi sample: dotted line with open diamonds).
}\label{fig-samp}
\end{center}
\end{figure}

Among 48 observed targets, 
  we use only 36 galaxies with median S/N per pixel 
  for the continuum greater than three for the following analysis.
Their basic information including the {\it IRAS} flux densities and 
  optical spectral properties is summarized in Table \ref{table1}.
We plot the infrared luminosity versus redshift of our sample in Figure \ref{fig-samp}.
The (U)LIRGs in \citet{ima08, ima10a}, hereafter the Imanishi sample\footnote{
  This sample contains ULIRGs from the {\it IRAS} 1 Jy sample \citep{kim98}
  and LIRGs from the bright galaxy sample 
  \citep[BGS;][]{soi87,san95} and from the revised BGS \citep[][]{san03}.},
  are also plotted for comparison.
The two samples are overall in a similar range of infrared luminosities,
  while our sample is more numerous than the Imanishi sample at 0.2 $< z <$ 0.4.
Imanishi et al. also preferentially selected non-Seyfert (U)LIRGs.
In the result, non-Seyfert fractions in these samples are large compared to 
  the fractions among their parent samples 
  (from $\sim$55\% to $\sim$70\% in ULIRGs and from $\sim$65\% to $\sim$80\% in LIRGs).

\section{OBSERVATIONS AND DATA REDUCTION}

We observed the target galaxies with the IRC spectrograph \citep{ona07} 
  on-board the {\it AKARI} satellite.
It is a slitless spectroscopy, but the targets are located inside a 1 $\times$ 1 arcmin$^2$ window
  in order to avoid the spectral overlap with nearby sources.
The 2.5--5 $\mu$m wavelength range was covered at a spectral resolution of R $\sim$120 
  ($\sim$2500 km s$^{-1}$) with NG grism. 
In the observing modes, the total on-source exposure time per pointing is about six minutes, 
  and the 3$\sigma$ detection limit is roughly 0.6 mJy.
Each target was observed one to eight times (mostly twice), depending on its visibility and
  on $K_s$-band magnitude.
The observation log is given in Table \ref{table2}.
Note that the NULIZ data were obtained in Phase 2 cooled by liquid helium,
  while the CLNSL and NISIG data were obtained in Phase 3 cooled by a cryocooler.

Data reduction was carried out using the {\small IDL} packages prepared by the 
  {\it AKARI} team: ``IRC Spectroscopy Toolkit Version 20101025" for the NULIZ data and 
  ``IRC Spectroscopy Toolkit for Phase 3 data Version 20101025" for the CLNSL and NISIG data.
Both packages are available online\footnote{ 
  http://www.ir.isas.jaxa.jp/ASTRO-F/Observation/DataReduction/IRC/}.
This involves dark subtraction, linearity correction, flat-fielding, 
  sub-frame combining, source detection, background subtraction and source extraction,
  wavelength calibration, and flux calibration.
Because one pointing consists of eight (or nine) sub-frames, 
  cosmic rays are efficiently removed. 
The wavelength calibration accuracy is $\sim$0.01 $\mu$m ($\sim$1 pixel).
The absolute flux calibration accuracy is $\sim$10\% at the central wavelength 
  and can be larger than 20\% at the edge \citep[][]{ohy07}.

We derived the spatial profile of each source in the two-dimensional spectra 
  obtained with the slitless spectrograph.
To extract one-dimensional spectra, 
  we used 3--9 pixels (3--9 $\times$ 1\arcsec.46) along the spatial direction.
The aperture width (see column 2 in Table 3) was determined to contain 
  signals larger than
  the background noise level, which is the root-mean-square (rms) of fluxes at 
  each side 11--15 pixel away from the center of the source.
Since the aperture is large enough to cover the NIR emission even for extended sources, 
  no aperture correction is applied.
For the sources observed several times, the spectra were stacked by taking the median.
The data collected during the late part of Phase 3 of {\it AKARI} are 
  sometimes seriously affected by bad pixels and background noise 
  because of the increase of the temperature in IRC.
If there were some data sets substantially inferior to others, 
  they were not used to obtain the final spectra.

\begin{figure}
\includegraphics[angle=0,scale=1.]{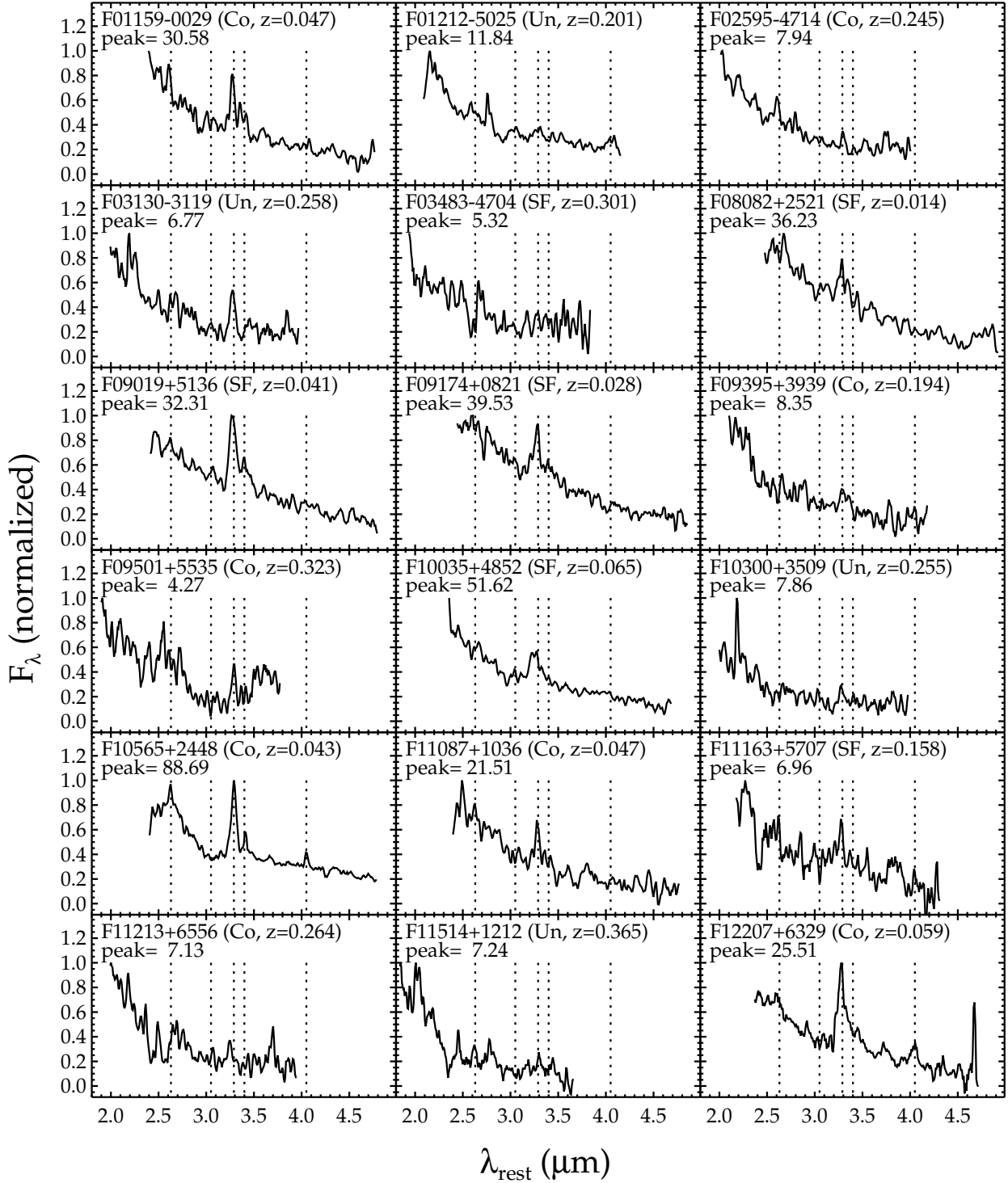}
\caption{
{\it AKARI} NIR spectra of 36 (U)LIRGs in our sample. 
For each object, their optical spectral type (SF=star-forming; Co=composite; LI=LINER; S2=Seyfert 2; S1=Seyfert 1; Un=Unclassified)
and redshift are shown.
The spectra presented here are smoothed using a 3 pixel box ($\sim$0.03 $\mu$m) and are normalized to peak values,
denoted in units of 10$^{-13}$ ergs s$^{-1}$ cm$^{-2}~\mu$m$^{-1}$.
The dotted lines indicate the wavelengths of Br$\beta$ (2.63 $\mu$m), H$_{2}$O ice-covered dust absorption (3.05 $\mu$m),
PAH emission (3.29 $\mu$m), bare carbonaceous dust absorption (3.4 $\mu$m), 
and Br$\alpha$ (4.05 $\mu$m) features.
}\label{fig-sp}
\end{figure}

We display the final spectra of 36 (U)LIRGs with the continuum S/N $>$ 3 
  in Figure \ref{fig-sp}.
These spectra show various continuum shapes, ranging 
  from steeply decreasing ones to nearly flat ones with increasing wavelength.
In most spectra, the PAH emission feature at 3.29 $\mu$m is clearly seen.
Br$\alpha$ (2.63 $\mu$m) and Br$\alpha$ (4.05 $\mu$m) hydrogen recombination lines 
  and broad absorption features 
  from H$_{2}$O ice (3.05 $\mu$m) and bare carbonaceous dust (3.4 $\mu$m) 
  are detected in some cases.

\section{MEASUREMENTS}

Figure \ref{fig-speg} shows 
  an example of spectral fitting to the NIR spectrum.
We fit each emission line of 
  the Br$\alpha$, Br$\beta$, and 3.3 $\mu$m PAH 
  with a single Gaussian function after subtracting 
  the local continuum defined by a linear fit around the line.
Especially for the 3.3 $\mu$m PAH profile, 
  we choose an asymmetric fitting range (3.15--3.35 $\mu$m) 
  in order to avoid contamination of the 3.4 $\mu$m sub-peak (see \citealt{ima10a}).
We compute the equivalent width of the PAH emission
  using the fitted Gaussian profile on top of the global continuum (see next paragraph).
By using the global continuum instead of the local continuum,
  we can reduce the effect of adjacent absorption feature on the equivalent width measurement. 
We set an upper limit for the non-detection by assuming that 
  the line has a Gaussian profile with a full width at half maximum (FWHM)
  and with an amplitude twice the local rms of the continuum-subtracted spectrum.
As for the fixed FWHMs, 
  we use the typical values measured from the {\it AKARI} spectra in this study:
  2500 km s$^{-1}$ for the Bracket lines and 4000 km s$^{-1}$ for the PAH feature.
The linewidth of the Bracket lines is well matched to the instrumental resolution.
The linewidth of the PAH feature is consistent with the PAH profile in other studies 
  (e.g., type-1 sources in \citealt{tok91}).
   
The global continuum slope $\Gamma$ is determined 
  using a single power-law model ($F_\lambda$ $\varpropto$ $\lambda^{\Gamma}$) with 
  data points less-affected by line features in the observed wavelengths at 2.7--4.8 $\mu$m.
The data points near the edge of the spectra are excluded 
  because of their large flux uncertainty \citep[][]{ohy07}.
The optical depths of the 3.1 and 3.4 $\mu$m dust absorption features 
  ($\tau _{3.1}$ and $\tau _{3.4}$, respectively) are calculated as 
  the natural logarithmic ratios between the adopted continuum levels 
  and the observed absorption profiles.
The continuum levels are adopted from the global continuum, and
  the absorption profiles are obtained after smoothing the spectrum with a 10 pixel box
  ($\sim$0.1 $\mu$m) to minimize noise effect.
We consider that the absorption feature is detected 
  when the maximum difference between the continuum level and the absorption profile 
  is twice larger than the local rms.

Measurements are listed in Table \ref{table3}.
The uncertainties estimated by considering the local rms values are also listed.
The median S/Ns of detected features are 6.4, 5.2, 3.6, 3.3, and 3.1 for 
  3.3 $\mu$m PAH, Br$\alpha$, Br$\beta$ fluxes, $\tau _{3.1}$, and $\tau _{3.4}$,
  respectively.  
We present the equivalent width of 3.3 $\mu$m PAH emission line (EW$_{\rm 3.3 PAH}$) 
  versus continuum slope ($\Gamma$) diagram in Figure \ref{fig-gamew}.
It shows that most (U)LIRGs have EW$_{\rm 3.3 PAH} <$ 150 nm and $-3 < \Gamma < 0$;
  the distribution of our sample is not different from that of the Imanishi sample.

\section{RESULTS AND DISCUSSION}

\subsection{AGN Diagnostics}

\subsubsection{AGN signature in the NIR spectra}\label{agnsig}

Emission at 3.3 $\mu$m is a prominent feature in (U)LIRG NIR spectra.
It probably originates from the reprocessing of ultraviolet (UV) radiation by PAH molecules.
The contribution of Pf$\delta$ emission line at a similar wavelength is usually negligible.
If (U)LIRGs host AGN, the EW$_{\rm 3.3 PAH}$ is suppressed 
  because hot dust emission from the AGN increases the NIR continuum flux level 
  and X-ray photons may destroy PAH molecules \citep[e.g.,][]{smi07}.
A red NIR continuum (i.e., a high value of continuum slope)
  as well as strong absorption features at 3.1 and 3.4 $\mu$m
  from H$_{2}$O ice-covered dust grains inside molecular clouds
  and bare carbonaceous dust in the diffuse interstellar medium, respectively 
  \citep[see][]{dra03}, indicate the presence of highly obscured compact sources.
While many of these absorbed sources with red continuum slopes 
  have been shown to harbor buried AGN \citep[e.g.,][]{ris06, san08, ima10a},
  deeply buried starbursts can also produce the same spectral signatures 
  \citep[e.g.,][]{des07, spo07, vei09a} and in general 
  all that is required is a warm, highly obscured heating source.
The small PAH equivalent width condition is useful for identifying weakly obscured AGN, 
  while the large continuum slope and optical depth conditions are efficient
  in detecting highly obscured AGN.
Therefore, all of these diagnostics are necessary to detect 
  as many obscured AGN as possible.

Following the criteria in \citet{ima10a}, we regard EW$_{\rm 3.3 PAH} <$ 40 nm,
  $\Gamma~(F_\lambda \varpropto \lambda^{\Gamma}) > -$1,
  $\tau _{3.1} >$ 0.3, and $\tau _{3.4} >$ 0.2 as AGN signatures,
  and classify sources satisfying at least one of these conditions 
  as NIR AGN-detected galaxies (see column 10 in Table \ref{table3}).
In the results, we find 19 AGN out of 36 (U)LIRGs
  based on these criteria.
Among these AGN, there are 13 sources with EW$_{\rm 3.3 PAH} <$ 40 nm (68\%), 
  five sources with $\Gamma > -$1 (26\%), 
  five sources with $\tau _{3.1} >$ 0.3 (26\%), and 
  three sources with $\tau _{3.4} >$ 0.2 (16\%).
Note that some sources satisfy more than one criterion.
Therefore, EW$_{\rm 3.3 PAH} <$ 40 nm is the primary criterion to select AGN.
A similar trend is also seen in the Imanishi sample 
  (56, 43, 45, and 10\% for 
  EW$_{\rm 3.3 PAH}$, $\Gamma$, $\tau _{3.1}$, and $\tau _{3.4}$, respectively).

We count the number of sources with AGN signatures 
  among our sample (U)LIRGs in bins of optical spectral type and of infrared luminosity.
These results are summarized in Table \ref{table4} together with those from the Imanishi sample.
The NIR AGN detection rates for our sample and the Imanishi sample are on average 53\% and 51\%,
  respectively, and agree well in each bin. 
\citet{ima08,ima10a} found that 
  the AGN signature in NIR spectra is more often detected in optical AGN-like 
  and in more luminous infrared galaxies.
Our sample appears to follow these trends, as shown in Figure \ref{fig-agndet}. 
However, it is not conclusive with our data alone 
  because of large uncertainties, in particular, 
  for the NIR AGN detection rate depending on optical spectral type.
In the combined sample of 180 (U)LIRGs with $L_{\rm IR} \geqq 10^{11} L_{\odot}$,
  the NIR AGN detection rate depends on optical spectral type as follows:
  36\% for SF (U)LIRGs, 55\% for composite (U)LIRGs, and 66\% for Seyfert 2 (U)LIRGs.
Note that most of previously classified LINERs are called composites in this study
  because we adopted the selection criteria of \citet{kew06} 
  rather than those of \citet{vei87} that were used in \citet{ima08,ima10a}.
There are two Seyfert 1 (U)LIRGs in the combined sample,
  and both of them show AGN signatures in their NIR spectra.
The total NIR AGN detection rates for LIRGs and ULIRGs are
  29\% and 65\%, respectively.
If we select (U)LIRGs without any priority to non-Seyfert galaxies,
  the NIR AGN detection rates for (U)LIRGs
  increase slightly [e.g., 30\% for LIRGs and 70\% for ULIRGs
  based on the (U)LIRG sample in \citet{yua10}].
 
When we consider non-Seyfert galaxies with AGN signature in the NIR
  to be optically elusive buried AGN,
  the buried AGN fraction (i.e., the number ratio of buried AGN to non-Seyferts)   
  increases with infrared luminosity: 
  24\% for (U)LIRGs with $L_{\rm IR} = 10^{11}$--$10^{12} L_{\odot}$, 
  60\% for (U)LIRGs with $L_{\rm IR} = 10^{12}$--$10^{12.3} L_{\odot}$, and 
  84\% for (U)LIRGs with $L_{\rm IR} = 10^{12.3}$--$10^{13} L_{\odot}$.
The higher AGN fraction in more luminous infrared galaxies has been reported 
  in previous studies 
  based on the data from not only {\it AKARI} 
  but also the {\it Infrared Space Observatory} 
  \citep[{\it ISO}; e.g.,][]{lut98,rig99,tra01} and 
  {\it Spitzer Space Telescope} \citep[e.g.,][]{des07,ima09a,val09,vei09a,pet11},
  suggesting an important role of AGN in increasing infrared luminosity.

As expected,
  we find many non-Seyfert (U)LIRGs with AGN signatures 
  in their NIR spectra [buried AGN; 49\% (55/113)].
On the other hand, a substantial number of Seyfert (U)LIRGs 
  do not show the AGN signatures in the NIR [33\% (14/43)].
The apertures used in the optical spectroscopy
  (our sample: 3\arcsec\ diameter fiber; Imanishi sample: 
  2\arcsec\ wide slit and 2 kpc extraction)
  are smaller than those of NIR spectroscopy (covers the entire galaxy size).
The line measurements in the optical spectra for our sample are aperture-corrected 
  following the method in \citet{hop03}, 
  but those for the Imanishi sample are not.  
The small aperture used for the optical spectral classification 
  may miss the extended emission associated with star formation, 
  and hence be more sensitive to weak, central AGN.
This helps to understand the existence of Seyferts without AGN signatures 
  in the NIR.
We find no difference in the redshift distribution
  between Seyferts with and without AGN signatures in the NIR spectra.
The different aperture size between the optical and NIR spectroscopy 
  seems partially responsible for the disagreement of spectral types.

Regardless of the aperture effect, 
  if the hot dust emission from AGN is very weak 
  because of a tiny covering of dust around the AGN,
  such AGN are clearly visible in the optical spectrum.
However, they would be not distinguishable from SF-dominated galaxies 
  based on the NIR diagnostics \citep[see][]{nar10}.
In contrast, when AGN are really heavily obscured,
  both optical and NIR diagnostics are less powerful.
Then the observations at other wavelengths are necessary 
  to detect them (e.g., X-ray: \citealt{bau10,ten10}; 
  mid/far-infrared: \citealt{far07,spo07,vei09a,hat10,elb11}; 
  but see also \citealt{elb10,hwa10b}; radio: \citealt{saj08,ima10b}).
  
In Figure \ref{fig-bpt}, we present the optical AGN diagnostic diagram based on 
  [O{\sc iii}]$\lambda5007$/H$\beta$ and [N{\sc ii}]$\lambda6584$/H$\alpha$ line ratios,
  and NIR AGN diagnostic diagram based on EW$_{\rm 3.3 PAH}$ and continuum slope.
In panel (a), the optical AGN (Seyfert$+$LINER) without the NIR AGN signature 
  have low [O{\sc iii}]/H$\beta$ ratios compared to those with AGN signatures 
  both in optical and in NIR spectra.
On the other hand, in panel (b), 
  the NIR properties of optical AGN without the NIR AGN signature 
  are not significantly different 
  from those of non-AGN both in optical and in NIR spectra.

\subsubsection{AGN contribution to the infrared luminosity}\label{agncont}

To measure the contribution of buried AGN to the infrared energy budget of (U)LIRGs,
  we use the infrared spectral energy distribution (SED) templates and fitting
  routine of \citet[][]{mul11}, {\small DECOMPIR}\footnote{http://sites.google.com/site/decompir}.
These templates consist of one AGN and five host-galaxy SEDs.
For the host-galaxy SEDs, {\it Spitzer} mid-infrared spectra of 
  starburst galaxies are extrapolated to the far-infrared using {\it IRAS} photometry.
These host-galaxy SEDs are grouped into five categories, 
  referred to as `SB1' through `SB5', in terms of 
  their overall shape and relative strength of PAH features.
For the AGN SED, the intrinsic SEDs of AGN-dominated sources are derived 
  after subtracting suitable host-galaxy components from the observed SEDs,  
  and these SEDs are averaged.
The infrared SEDs of AGN show a large spread, mainly dependent 
  on dust distribution around AGN (i.e., smooth vs. clumpy torus structures).
However, the different AGN SEDs do not significantly change the resulting 
  AGN contribution to the infrared luminosity in (U)LIRGs 
  \citep[see][]{mul11,poz12}.
Based on this SED fitting with sparse photometric data points such as {\it IRAS},  
  \citet[][]{mul11} found that the intrinsic AGN luminosities measured
  are actually correlated with those 
  from high-resolution mid-infrared observations of the AGN cores.

We apply this routine to 69 (U)LIRGs with S/Ns $>$ 3 at all four {\it IRAS} bands 
  (12, 25, 60, and 100 $\mu$m) in the combined sample, and fit their SEDs five times 
  with AGN and one of host-galaxies
  by allowing renormalization of these two templates.
We choose the best-fit solution with the lowest $\chi^2$ value, 
  computing the AGN contribution to the infrared (8--1000 $\mu$m) luminosity 
  from this template set.
Figure \ref{fig-sed} represents example SEDs
  with the best-fit AGN and host-galaxy templates.
The AGN contribution in (U)LIRGs ranges from 0\% to 69\%,
  and is on average 6--8\% in LIRGs and 11--19\% in ULIRGs.
Because the combined sample preferentially includes non-Seyferts, 
  the AGN contribution in this study seems to be small compared to other studies 
  (e.g., LIRGs: $\sim$10\% in \citealt{pet11}; 
  ULIRGs: $\sim$20\% in \citealt{far03}; 35--40\% in \citealt{vei09a}; 
  $\sim$25\% in \citealt{nar10}; 15--20\% in \citealt{ris10}).

Figure \ref{fig-agncont} shows
  the correlation of the AGN contribution with 
  the presence of AGN signature in the NIR, optical spectral type, and infrared luminosity.
Not surprisingly,
  the AGN contribution is higher 
  in (U)LIRGs with AGN signature in the NIR
  than in those without AGN signature.
The AGN contribution
  clearly increases with increasing infrared luminosity \citep[see also][]{don12},
  similar to the trend of buried AGN fraction in Figure \ref{fig-agndet} (b).
The AGN contribution for Seyfert (U)LIRGs
  is slightly larger than those for SF and composite (U)LIRGs.

\subsubsection{AGN diagnostics with {\it WISE} data}\label{agnwise}  

Recently, the {\it Wide-field Infrared Survey Explorer} \citep[{\it WISE};][]{wri10}
  opens up the opportunity to probe mid-infrared properties 
  (3.4, 4.6, 12, and 22 $\mu$m) for a large sample of galaxies 
  with excellent sensitivity.
We use the {\it WISE} all-sky survey source catalog\footnote{
  http://wise2.ipac.caltech.edu/docs/release/allsky/}
  to identify the {\it WISE} counterparts of the (U)LIRGs observed with {\it AKARI} IRC
  and (U)LIRGs at 0.01 $< z <$ 0.4 in the SDSS DR7 \citep{hwa10a}
  within 3\arcsec .

In Figure \ref{fig-wise} (a), 
  we compare the continuum slope $\Gamma$ with {\it WISE} [3.4]--[4.6] color 
  \citep[Vega magnitude system;][]{jar11} for the {\it AKARI} (U)LIRGs.
We overplot the expected {\it WISE} [3.4]--[4.6] colors 
  from simple power-law continuum models
  as a function of continuum slope (solid line).
The continuum slope $\Gamma$ and {\it WISE} [3.4]--[4.6] color 
  show a tight correlation 
  (Spearman rank correlation coefficient = 0.82;
  the probability of obtaining the correlation by chance = 1.03$\times10^{-30}$)
  with some offset and scatter
  around the expected relation.
Most sources have small values of continuum slope compared to the expectation
  because the presented slopes (fitting range: 2.7--4.8 $\mu$m) are 
  affected by the spectrum at $<$3.4 $\mu$m where
  the stellar population contribution is large (see \citealt{lee10}).  
The scatter may come from the contamination by the 3.3 $\mu$m PAH feature.
For the galaxies at $z <$ 0.13, the presence of 3.3 $\mu$m PAH emission makes 
  [3.4]--[4.6] color bluer than the color without the PAH emission.
On the other hand, if the galaxies are at 0.26 $< z <$ 0.55,
  the presence of 3.3 $\mu$m PAH emission makes [3.4]--[4.6] color redder than 
  the color without the PAH emission.
The amount of change in colors depends on redshift and on the strength of PAH emission.
From the experiment with the spectra of our sample,
  we find that the presence of PAH emission can change the [3.4]--[4.6] color 
  by $\pm$0.2 mag, consistent with the scatter in Figure 9 (a).

We find that the AGN selection criterion based on the continuum slope (i.e., $\Gamma$ $> -$1)
  in this study is roughly equivalent to that of [3.4]--[4.6] $> 0.8$ suggested by \citet{ste12}
  \citep[see also][]{ass10, jar11}.
If (U)LIRGs with AGN signatures in the NIR spectra are regarded as genuine AGN (filled symbols), 
  the {\it WISE} color criterion selects AGN with 
  72\% completeness (51 out of 71 genuine AGN satisfy the {\it WISE} color criterion) and
  76\% reliability (51 out of 67 objects which satisfy the {\it WISE} color criterion are genuine AGN).

Figure \ref{fig-wise} (b) shows the 
  {\it WISE} [3.4]--[4.6] colors  versus 
  {\it IRAS} flux density ratios between 25 and 60 $\mu$m 
  (hereafter {\it IRAS} 25--60 $\mu$m colors) for the SDSS (U)LIRG sample.
The {\it IRAS} 25--60 $\mu$m color
  is known to be associated with nuclei activity 
  in infrared-luminous galaxies \citep[e.g.,][]{deg85,san88b,nef92,vei09a,lee11}.
AGN-dominated galaxies show warm {\it IRAS} 25--60 $\mu$m colors ($f_{25}/f_{60} \geqq 0.2$),
  while SF-dominated galaxies show cool colors ($f_{25}/f_{60} < 0.2$).
The composite and SF galaxies show similar distributions in this domain,
  but the AGN are significantly different.
The median colors of AGN differ from those of non-AGN with significance levels 
  of 6.8$\sigma$ and 7.4$\sigma$ in the {\it IRAS} and {\it WISE}, respectively.
Interestingly, there are a substantial number of SF (U)LIRGs 
  in the lower-right corner.
It seems real
  even if we consider large uncertainties associated with the {\it IRAS} colors.
They have warm dust emission without hot dust emission,
  therefore appearing to be heavily obscured AGN. 
As a result,
  the {\it WISE} [3.4]--[4.6] color is a good tracer of AGN-heated hot dust emission,
  but may not be sufficient to detect heavily obscured AGN.

\subsection{Comparison between Optical and Infrared Properties}\label{comp} 

\subsubsection{Star formation rate indicators}\label{sfr}

The total infrared continuum, 3.3 $\mu$m PAH emission, 
  and recombination lines including H$\alpha$ and Br$\alpha$ of galaxies
  are useful indicators of star formation rate (SFR) \citep[][]{ken98}.
In Figure \ref{fig-sfr} (a--b), we compare these SFR indicators: 
  infrared luminosity versus (a) H$\alpha$ and (b) 3.3 $\mu$m PAH luminosities.
The H$\alpha$ luminosities are extinction-corrected using the Balmer decrement and
  \citet{cal00} extinction curve.
The expected relationships are overplotted between these parameters 
  (dotted lines; hereafter SF galaxy sequences):
  $L_{\rm H\alpha}/L_{\rm IR}=10^{-2.25}$ from the empirical relation in \citet{ken98} and 
  $L_{\rm 3.3 PAH}/L_{\rm IR}=10^{-3}$ from the observations of \citet{mou90} and \citet{ima02}.
These panels show that 
  there are large offsets and scatters between the data and the expected relationships, 
  even for SF (U)LIRGs without any AGN signature [pure SF (U)LIRGs; large filled symbols].
The offset from the SF galaxy sequence is larger in ULIRGs than in LIRGs,
  suggesting that H$\alpha$ and 3.3 $\mu$m PAH emissions 
  are more depressed in ULIRGs.

There could be several reasons
  for the strong depression of H$\alpha$ and 3.3 $\mu$m PAH emission in ULIRGs.
(1) The amount of dust extinction could be systematically underestimated in ULIRGs.
To check this effect, we plot the observed line ratios of 
  Br$\alpha$/H$\alpha$ and H$\alpha$/H$\beta$ in panel (c).
The pure SF (U)LIRGs appear to have larger Br$\alpha$/H$\alpha$ ratios by considering 
  the extinction curve of \citet{cal00} with a typical $R_{\rm V}$ value of 4.05.
This can imply the need of a larger $R_{\rm V}$ value for the extinction correction 
  in (U)LIRGs [e.g., $R_{\rm V}$ = 5.01 denoted by the arrow].
Some studies also suggest the need of flatter/grayer extinction curves 
  (i.e., large $R_{\rm V}$ values) in ULIRGs,
  which may be attributed to supernovae-driven large-size dust grains
  (e.g., \citealt{kaw11}; \citealt{shi11}; see also \citealt{boq12}).   
However, because of large scatter in the data and different aperture size
  between the optical and NIR spectroscopy,
  it should be checked with a more extensive data set in future studies.
(2) Even for ULIRGs without any AGN signature in optical and NIR observations, 
  there could still be hidden AGN that play a role in the relative depression of line emission.
To remove contamination of hidden AGN, 
  it is necessary to search for AGN at other wavelengths.
However, it is expected that the contribution of buried AGN is not 
  significant as discussed in Section \ref{agncont}.
(3) The line emission from ULIRGs may be intrinsically weak.
Regardless of the presence of AGN, the star formation in ULIRGs 
  produces large infrared continuum emission
  because a larger fraction of stellar UV photons is absorbed by dust inside 
  star-forming regions under the strong radiation field in ULIRGs \citep[][]{abe09}.
The PAH emission in ULIRGs may be depressed in the sense that 
  intense radiation fields do not produce photo-dissociation regions
  that are necessary for PAH emission, 
  or lead to the destruction of PAH carriers \citep[see][]{voi92,smi07,vei09a}.
It is still unclear whether the ionization state of grains is actually related to 
  infrared luminosity \citep[e.g.,][]{des07,smi07,vei09a,ima10a,pet11}.

\subsubsection{Dust extinction as a function of infrared color}\label{extinction}

In Figure \ref{fig-av}, we present the Br$\alpha$/H$\beta$ line ratio of (U)LIRGs 
  as a function of {\it IRAS} 25--60 $\mu$m color.
There is an anti-correlation between the two parameters with 
  Spearman rank correlation coefficient = $-$0.67 and
  the probability of obtaining the correlation by chance = 0.02.
This supports that warmer sources are less extinguished than cooler sources
  \citep[see][]{vei99a,vei09a,kee05,ima08}.
When using Balmer decrement values instead, such a correlation becomes weaker 
  since the optical lines do not trace well the buried sources.

\section{SUMMARY}

We obtained {\it AKARI} 2.5--5 $\mu$m spectra of 36 (U)LIRGs, 
  selected mainly from the {\it IRAS}-detected galaxies in the SDSS.
We measured the NIR spectral features including
 continuum slope, 3.3 $\mu$m PAH strength, 
 optical depths at 3.1 and 3.4 $\mu$m, and
 Bracket lines in order to find AGN signatures.
Together with samples in the literature and ancillary data,
  we compared optical and infrared properties of (U)LIRGs.
Our primary results are summarized below.

\begin{enumerate}
\item 
We found that
  52\% (14/27) of optical non-Seyfert galaxies in our (U)LIRG sample
  show the AGN signature in their NIR spectra.
The NIR AGN detection rate for the combined sample
  is higher in composite (U)LIRGs than in SF (U)LIRGs,
  and increases with infrared luminosity.
  
\item 
We fitted the {\it IRAS} photometric data 
  for 69 (U)LIRGs with the AGN/starburst SED templates 
  to compute the AGN contribution to the infrared luminosity.
We found that the contribution of buried AGN 
  to the infrared luminosity   
  is 5--10\%, smaller than the typical AGN contribution
  in (U)LIRGs including Seyfert galaxies (10--40\%).

\item 
The NIR continuum slopes correlate well with 
  {\it WISE} [3.4]--[4.6] colors.
Using the {\it WISE} [3.4]--[4.6] color vs. {\it IRAS} 25--60 $\mu$m color domain,
  we found sources associated with warm dust emission, but without hot dust emission.
The {\it WISE} color is useful for identifying a large number of AGN, 
  while it can miss heavily obscured AGN.

\end{enumerate}

\acknowledgments
We are grateful to an anonymous referee for his/her comments that helped  
to improve the manuscript.
This work was supported by Mid-career Research Program 
through NRF grant funded by the MEST (No.2010-0013875).
J.C.L., M.K., and J.H.L. are members of the Dedicated Researchers 
for Extragalactic AstronoMy (DREAM)
in Korea Astronomy and Space Science Institute (KASI).
H.S.H. acknowledges the Centre National d`Etudes Spatiales (CNES) 
and the Smithsonian Institution for the support of his post-doctoral fellowship.
This research is based on observations with {\it AKARI}, a JAXA project 
with the participation of ESA.
Funding for the SDSS and SDSS-II has been provided by the Alfred P. Sloan Foundation, 
the Participating Institutions, the National Science Foundation, 
the U.S. Department of Energy, the National Aeronautics and Space Administration, 
the Japanese Monbukagakusho, the Max Planck Society, 
and the Higher Education Funding Council for England.
The SDSS Web Site is http://www.sdss.org/.
The SDSS is managed by the Astrophysical Research Consortium for the Participating Institutions. 
The Participating Institutions are the American Museum of Natural History, 
Astrophysical Institute Potsdam, University of Basel, University of Cambridge, 
Case Western Reserve University, University of Chicago, Drexel University, Fermilab, 
the Institute for Advanced Study, the Japan Participation Group, Johns Hopkins University, 
the Joint Institute for Nuclear Astrophysics, the Kavli Institute for Particle Astrophysics 
and Cosmology, 
the Korean Scientist Group, the Chinese Academy of Sciences (LAMOST), 
Los Alamos National Laboratory, the Max-Planck-Institute for Astronomy (MPIA), 
the Max-Planck-Institute for Astrophysics (MPA), New Mexico State University, 
Ohio State University, University of Pittsburgh, University of Portsmouth, 
Princeton University, the United States Naval Observatory, and the University of Washington.
This publication makes use of data products from the {\it Wide-field Infrared Survey Explorer}, 
which is a joint project of the University of California, Los Angeles, 
and the Jet Propulsion Laboratory, California Institute of Technology, 
funded by the National Aeronautics and Space Administration.

\clearpage

\clearpage

\clearpage
\begin{figure}
\includegraphics[angle=0,scale=1.]{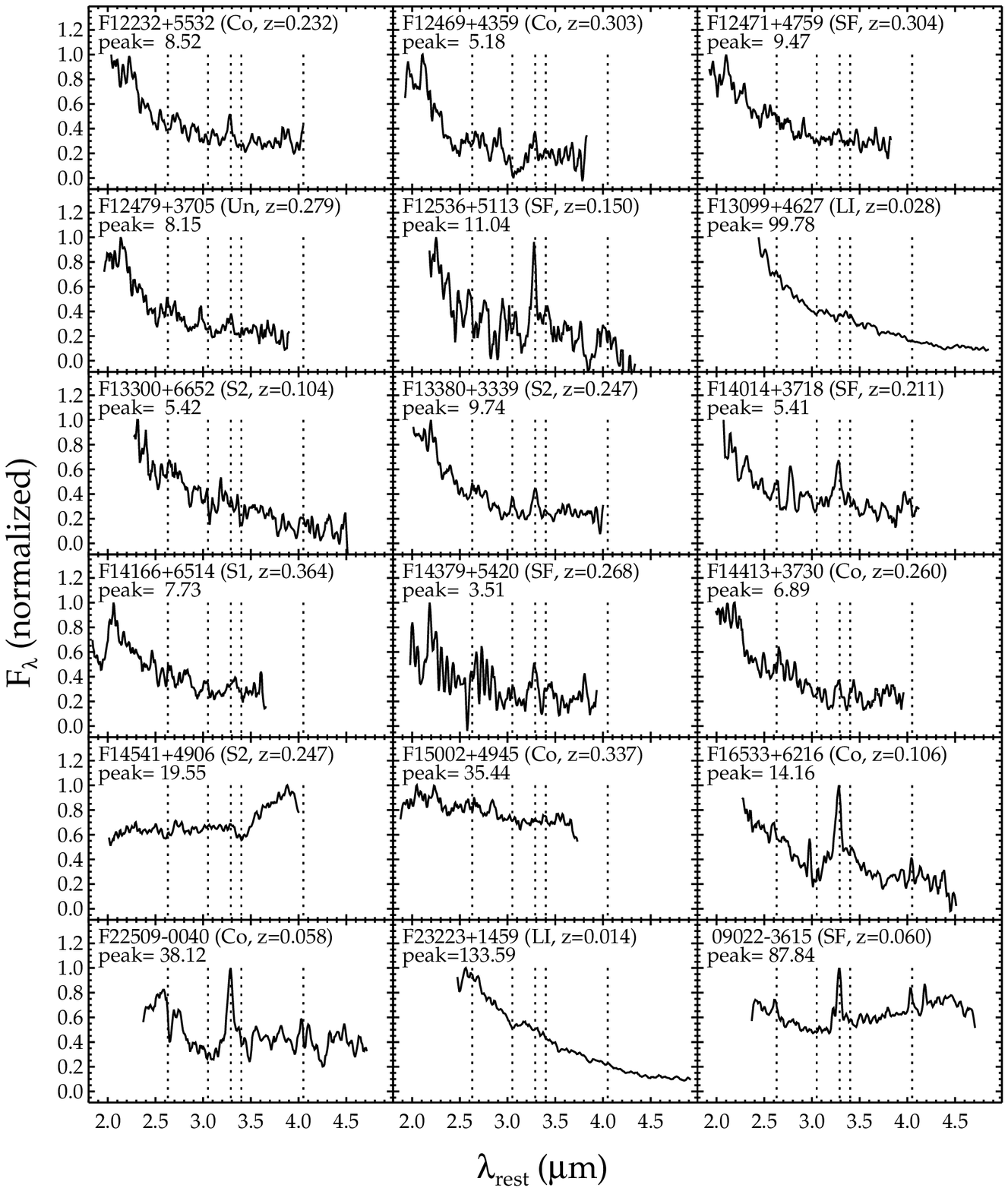}
\begin{center}
Fig. \ref{fig-sp}.--- Continued
\end{center}
\end{figure}

\clearpage
\begin{figure}
\begin{center}
\includegraphics[angle=0,scale=1.]{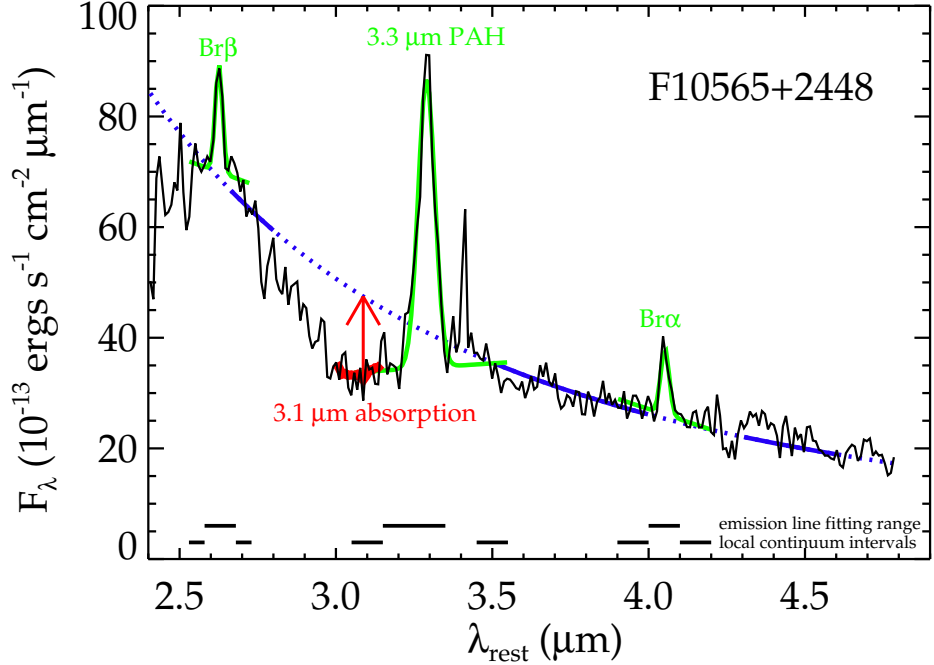}
\caption{
Example of spectral fitting for F10565$+$2448.
The background solid line is the observed spectrum.
The Gaussian fit lines to measure 2.63 $\mu$m Br$\beta$, 3.3 $\mu$m PAH, and 4.05 $\mu$m Br$\alpha$ emissions are overplotted.
The solid-dotted power-law line represents the global continuum.
The wavelength intervals not used in the continuum fit are denoted by dotted lines. 
The thick line around 3.1 $\mu$m is the 10 pixel smoothed profile of 3.1 $\mu$m absorption.
The arrow shows the wavelength where the difference 
  between the continuum level and absorption profile is maximized.
In this case, 3.4 $\mu$m absorption is not detected.
}\label{fig-speg}
\end{center}
\end{figure}

\clearpage
\begin{figure}
\begin{center}
\includegraphics[angle=0,scale=1.]{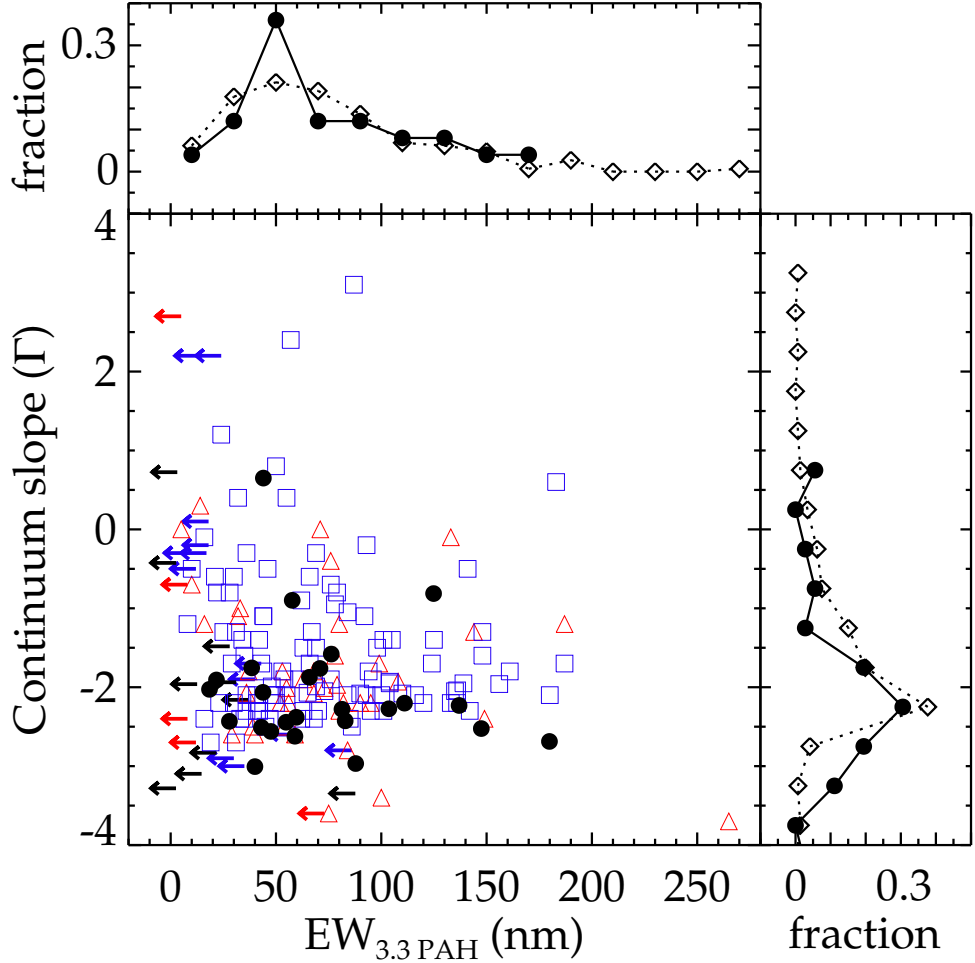}
\caption{
EW$_{\rm 3.3 PAH}$ vs. continuum slope $\Gamma$ diagram.
The (U)LIRGs in this study, \citet{ima08}, and \citet{ima10a} are represented 
by black filled circles, red open triangles, and blue open squares, respectively.
The sources with the upper limits of EW$_{\rm 3.3 PAH}$ are represented by arrows.
The continuum slope $\Gamma$ was defined by $F_\lambda$ $\varpropto$ $\lambda^{\Gamma}$ in this study, 
while it was defined by $F_\nu$ $\varpropto$ $\lambda^{\Gamma}$ in Imanishi et al.
Thus, $\Gamma_{\rm this~study}$ equals $\Gamma_{\rm Imanishi} - 2$, and
the values of $\Gamma_{\rm this~study}$ and $\Gamma_{\rm Imanishi} - 2$ are presented for fair comparison.
The EW$_{\rm 3.3 PAH}$ and continuum slope distributions are shown in the upper and right panels, respectively 
(our sample: solid line with filled circles; Imanishi sample: dotted line with open diamonds).
In the upper panel, the (U)LIRGs with EW$_{\rm 3.3 PAH}$ upper limits are not included. 
}\label{fig-gamew}
\end{center}
\end{figure}

\clearpage
\begin{figure}
\begin{center}
\includegraphics[angle=0,scale=1.]{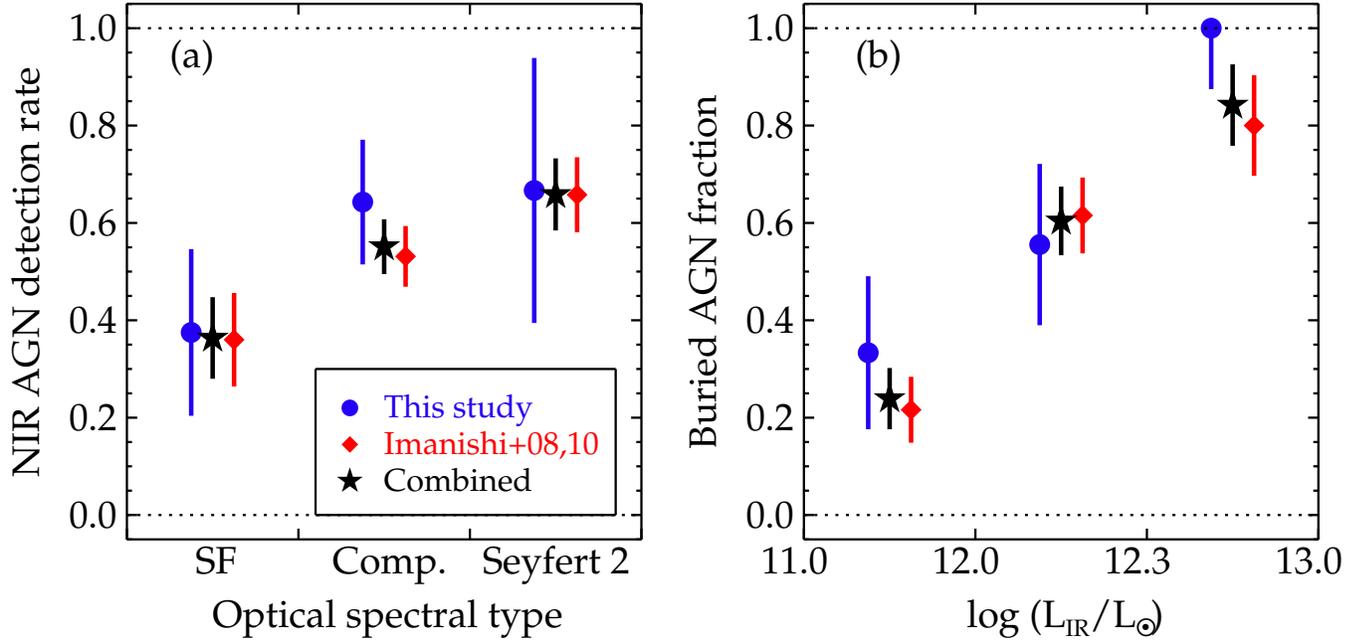}
\caption{
(a) {\it AKARI}-based AGN detection rate as a function of optical spectral type 
(See Section \ref{agnsig} for the definition of AGN signatures in the NIR spectra).
(b) Buried AGN fraction as a function of infrared luminosity.
The buried AGN fraction means the number fraction of NIR AGN among optical non-Seyfert galaxies.
The blue circles, red diamonds, and black stars indicate 
  the results from our sample, Imanishi sample, 
  and the combined sample, respectively.
The error bars are based on Poisson statistics \citep[see][]{dep04}.  
}\label{fig-agndet}
\end{center}
\end{figure}

\clearpage
\begin{figure}
\begin{center}
\includegraphics[angle=0,scale=1.]{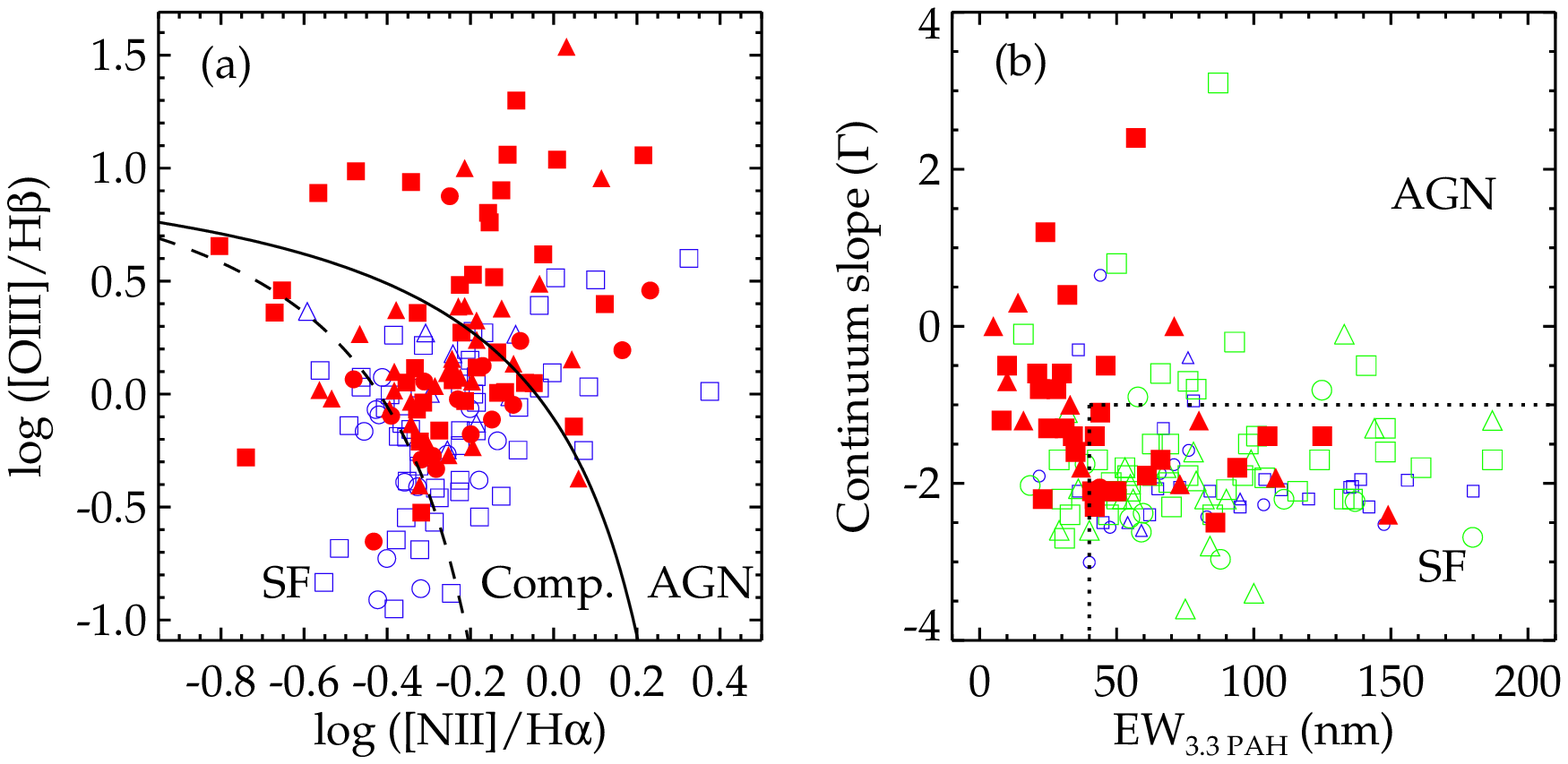}
\caption{
(a) Optical diagnostic diagram based on 
[O{\sc iii}]$\lambda5007$/H$\beta$ and [N{\sc ii}]$\lambda6584$/H$\alpha$ line ratios.
The (U)LIRGs in this study, \citet{ima08}, and \citet{ima10a} are represented 
by circles, triangles, and squares, respectively.
The optical line ratios of the Imanishi sample are taken from \citet{vei95,vei99a} and \citet{kew01}. 
The sources with (/without) AGN signatures in their NIR spectra are denoted by red filled 
(/blue open) symbols.
The solid and dashed lines indicate the maximum starburst \citep{kew01} and pure star formation \citep{kau03b} lines, respectively.  
(b) NIR diagnostic diagram based on EW$_{\rm 3.3 PAH}$ and continuum slope $\Gamma$.
The optical SF, composite, and AGN (LINER $+$ Seyfert) sources are denoted by small blue open, large green open, and large red filled symbols, respectively.
The (U)LIRGs optically unclassified or with upper limits are not included.
The dotted line shows the AGN selection criteria in this study: EW$_{\rm 3.3 PAH}<$ 40 nm or $\Gamma > -$1.
}\label{fig-bpt}
\end{center}
\end{figure}

\clearpage
\begin{figure}
\begin{center}
\includegraphics[angle=0,scale=1.]{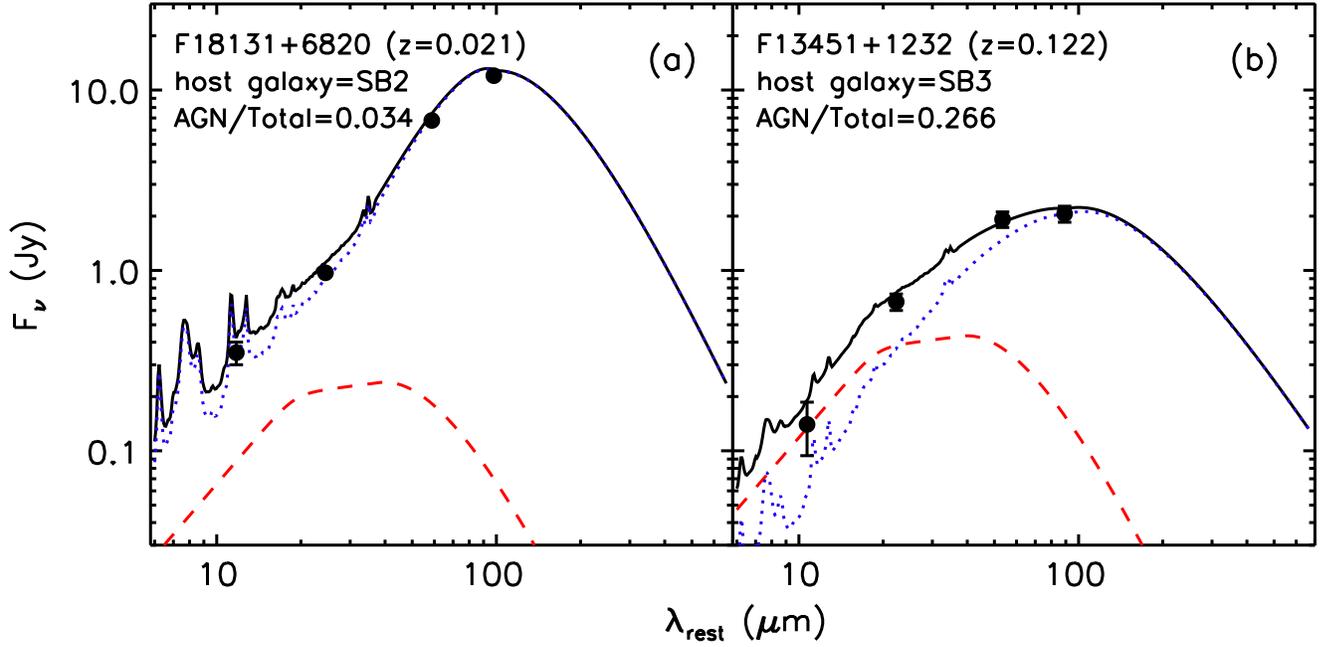}
\caption{
The best fit SEDs for (a) F18131+6820 and (b) F13451+1232 
  derived using the infrared SED-fitting routine, {\small DECOMPIR} \citep{mul11}.
The filled circles with errorbars indicate their four {\it IRAS} flux densities.
The black solid, blue dotted, and red dashed lines represent the total, host-galaxy, and AGN SEDs, respectively.
The labels in the upper left indicate
  which host-galaxy template is used and the derived AGN contribution to the total infrared luminosity.
}\label{fig-sed}
\end{center}
\end{figure}

\clearpage
\begin{figure}
\begin{center}
\includegraphics[angle=0,scale=1.]{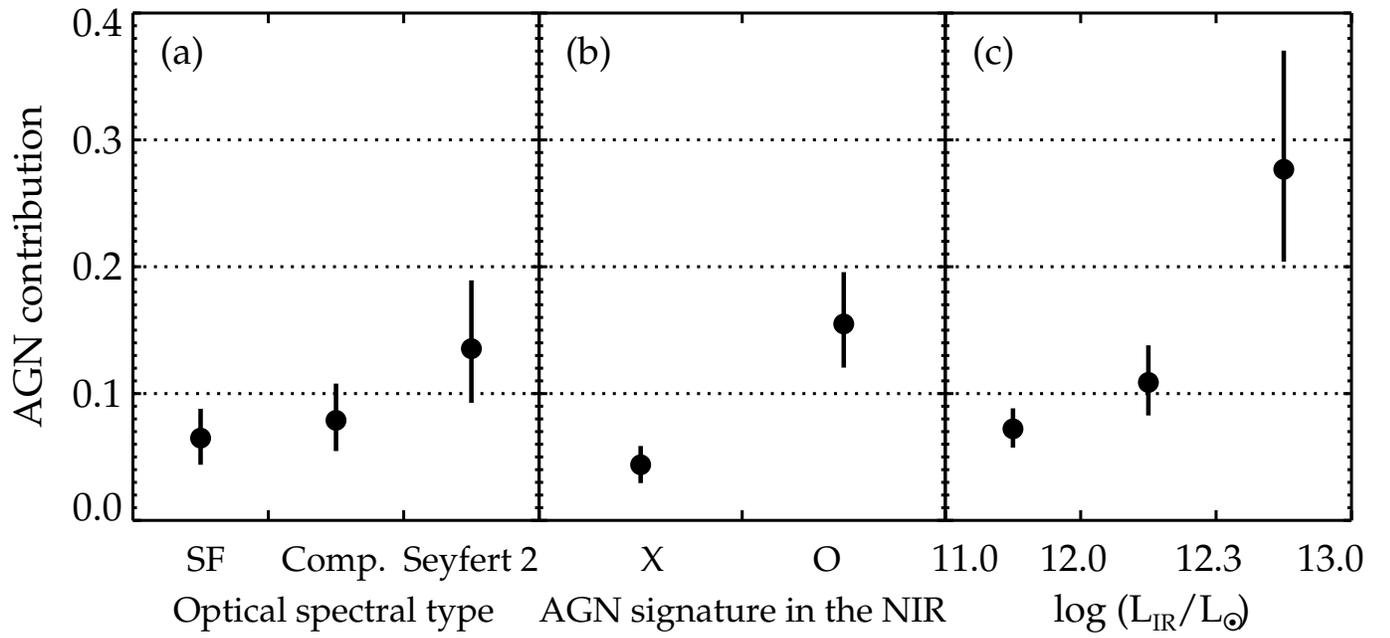}
\caption{
The AGN contribution to the infrared luminosities of (U)LIRGs in bins of 
(a) optical spectral type, (b) presence of AGN signature in the NIR spectra,
and (c) infrared luminosity.
The mean values with sampling errors are shown.
}\label{fig-agncont}
\end{center}
\end{figure}

\clearpage
\begin{figure}
\begin{center}
\includegraphics[angle=0,scale=1.]{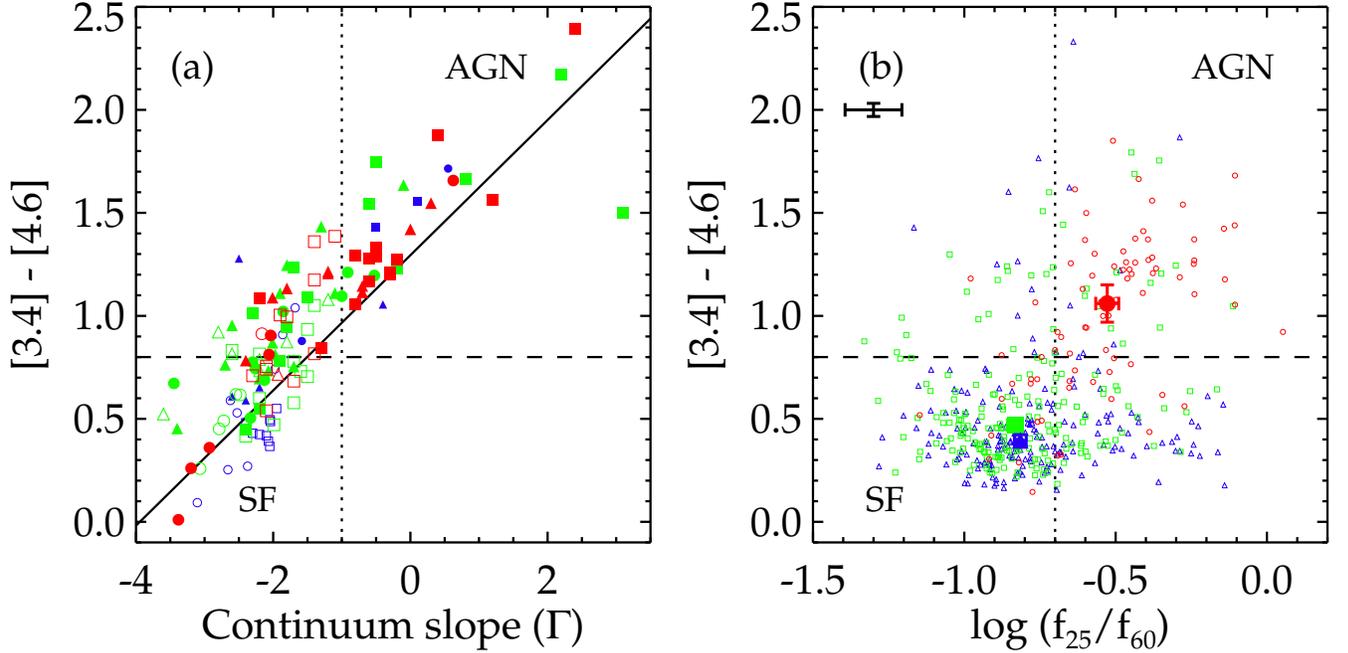}
\caption{
(a) Comparison between the {\it WISE} [3.4]--[4.6] color (Vega) and NIR continuum slope $\Gamma$ 
for (U)LIRGS observed with {\it AKARI} IRC.
The (U)LIRGs in this study, \citet{ima08}, and \citet{ima10a} are represented
by circles, triangles, and squares, respectively.
The optical composite and AGN (/SF) sources are denoted by large (/small) symbols,
and the sources with (/without) AGN signature in NIR spectra are shown by filled (/open) symbols.
The symbols are also color-coded according to optical spectral types
(blue: SF; green: composite; red: AGN).
The solid line represents the {\it WISE} colors expected from 
simple power-law continuum models ($F_\lambda$ $\varpropto$ $\lambda^{\Gamma}$).
The vertical dotted and horizontal dashed lines
indicate the criteria separating AGN from SF-dominated galaxies in this study 
and \citet{ste12}, respectively.
(b) The {\it WISE} [3.4]--[4.6] color vs. {\it IRAS} flux density ratio of 25 to 60 $\mu$m diagram.
The small open symbols are (U)LIRGs in SDSS DR7 \citep{hwa10a}.
Their median values are denoted by large filled symbols with sampling errors.
The red circles, green squares, and blue triangles are for AGN, composite and SF galaxies
determined in the optical spectra.
The (U)LIRGs with S/Ns $>$ 3 at ({\it WISE}) 3.4, 4.6, ({\it IRAS}) 25, 60 $\mu$m are only presented.
The error bars in the upper-left corner indicate the typical uncertainties
associated with the colors.
The vertical dotted and horizontal dashed lines indicate 
the criteria separating AGN from SF-dominated galaxies in \citet{san88b} and \citet{ste12}, respectively.
}\label{fig-wise}
\end{center}
\end{figure}

\clearpage
\begin{figure}
\begin{center}
\includegraphics[angle=0,scale=1.]{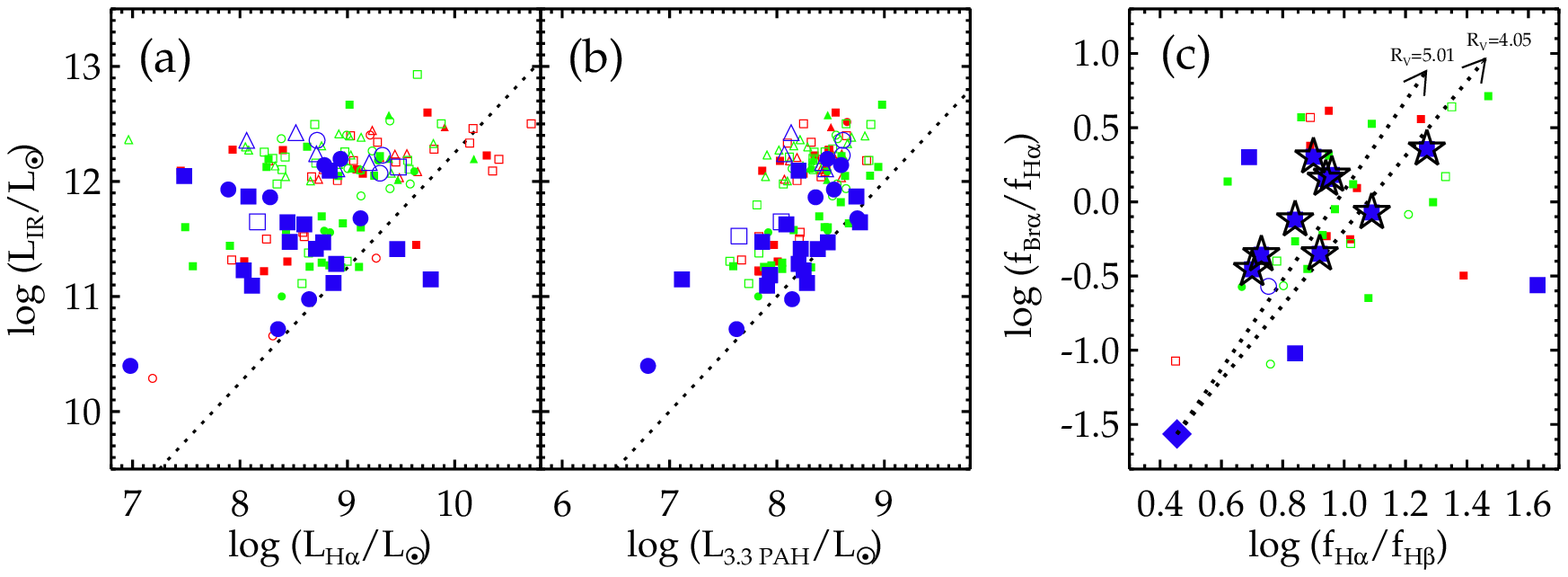}
\caption{Comparisons of infrared luminosity with (a) H$\alpha$ and (b) 3.3 $\mu$m PAH
luminosities.
The (U)LIRGs in this study, \citet{ima08}, and \citet{ima10a} are represented
by circles, triangles, and squares, respectively.
The optical SF (/composite and AGN) sources are denoted by large (/small) symbols, 
and the sources without (/with) AGN signature in NIR spectra are shown by filled (/open) symbols.
The symbols are also color-coded according to optical spectral types
(blue: SF; green: composite; red: AGN).
In each panel, the dotted line indicates the SF galaxy sequence (see Section \ref{sfr}).
(c) Comparison between the observed Br$\alpha$/H$\alpha$ and H$\alpha$/H$\beta$ line ratios.
The symbols are the same as in panels (a--b), but 
the diamond in the lower-left corner indicates the intrinsic position of SF galaxies in this plane 
(case B theory; \citealt{ost06}).
The reddening vectors of $A_{\rm V} = 8$ mag from the \citet{cal00} extinction curve
with $R_{\rm V} =$ 4.05 and 5.01 are shown by dotted line arrows.
Since the curve is given at shorter wavelengths than 2.2 $\mu$m,
the extinction at 4.05 $\mu$m ($A_{\rm Br\alpha}$) is extrapolated from $A_{\rm 2.2\mu m}$ using the model of 
$A_\lambda$ $\varpropto$ $\lambda^{-1.75}$ (e.g., \citealt{car89,dra89}; but see also \citealt{nis09}).
We fit a linear relation to the data of pure SF (U)LIRGs after removing three outliers 
with 2$\sigma$ clipping (open star symbols), and so obtain the curve with $R_{\rm V} =$ 5.01.
}\label{fig-sfr}
\end{center}
\end{figure}

\clearpage
\begin{figure}
\begin{center}
\includegraphics[angle=0,scale=1.]{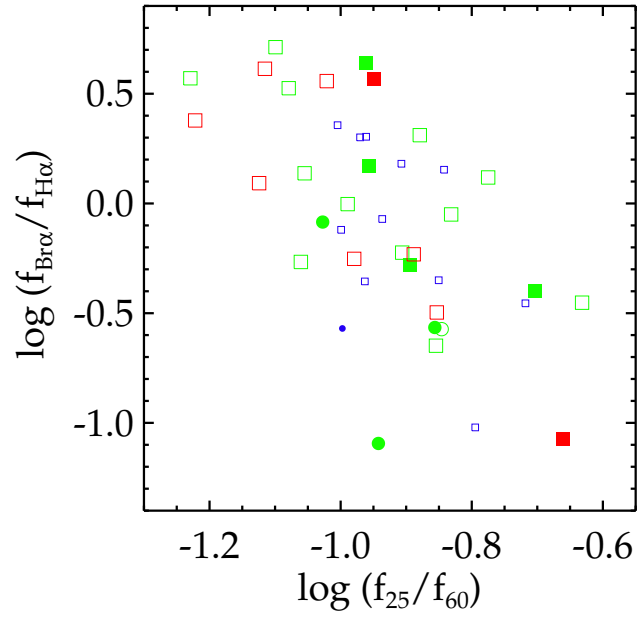}
\caption{The Br$\alpha$/H$\alpha$ line ratio vs. 
{\it IRAS} flux density ratio between 25 and 60 $\mu$m for the {\it AKARI} (U)LIRG sample.
The symbols are the same as in Figure 9 (a).
}\label{fig-av}
\end{center}
\end{figure}

\clearpage
\begin{deluxetable}{rcccrrrrccccccl}
\rotate
\tabletypesize{\scriptsize}
\tablecaption{Basic information and optical spectral properties.\label{table1}}
\tablewidth{0pt}
\tablehead{
\colhead{Object} &
\colhead{RA} &
\colhead{Dec} &
\colhead{z} &
\colhead{$f_{12}$}  &
\colhead{$f_{25}$}  &
\colhead{$f_{60}$}  &
\colhead{$f_{100}$}  &
\colhead{log $L_{\rm IR}$} &
\colhead{Type} &
\colhead{H$\alpha$/H$\beta$}&
\colhead{log $L_{\rm H\alpha}$} &
\colhead{[\mbox{O{\sc iii}}]/H$\beta$}&
\colhead{[\mbox{N{\sc ii}}]/H$\alpha$}&
\colhead{Ref.} \\
\colhead{(1)} & (2) & (3) & (4) & \colhead{(5)} & \colhead{(6)} & \colhead{(7)} & \colhead{(8)} & (9) & (10) & (11) & (12) & (13) & (14) & (15)
}
\startdata
  F01159$-$0029  &  01 18 34.1  &  $-$00 13 42  &  0.047  &  $<$0.222  &     0.335  &   3.476  &     5.013  &  11.56  &  Co  &    6.94  &    8.78  & 0.42 & 0.66 & ~~1/2  \\
  F01212$-$5025  &  01 23 19.9  &  $-$50 09 30  &  0.201  &  $<$0.056  &  $<$0.077  &   0.415  &     0.717  &  12.03  &  Un  &    ...   &    ...   & ...  & ...  & ~~3    \\
  F02595$-$4714  &  03 01 17.6  &  $-$47 02 14  &  0.245  &  $<$0.090  &  $<$0.063  &   0.268  &     0.484  &  12.11  &  Co  &    5.69  &    8.94  & 1.33 & 0.68 & ~~3/4  \\
  F03130$-$3119  &  03 15 03.5  &  $-$31 07 48  &  0.258  &  $<$0.087  &  $<$0.103  &   0.252  &     0.438  &  12.15  &  Un  &    ...   &    ...   & ...  & ...  & ~~3    \\
  F03483$-$4704  &  03 49 55.3  &  $-$46 55 20  &  0.301  &  $<$0.083  &  $<$0.098  &   0.156  &  $<$0.542  &  12.07  &  SF  &    4.78  &    9.24  & 0.80 & 0.41 & ~~3/4  \\
  F08082$+$2521  &  08 11 13.5  &  $+$25 12 24  &  0.014  &     0.124  &     0.251  &   2.243  &     6.116  &  10.40  &  SF  &    5.39  &    6.92  & 0.12 & 0.38 & ~~1/2  \\
  F09019$+$5136  &  09 05 28.9  &  $+$51 24 49  &  0.041  &     0.096  &     0.113  &   0.946  &     1.683  &  10.98  &  SF  &    5.27  &    8.58  & 0.19 & 0.40 & ~~1/2  \\
  F09174$+$0821  &  09 20 08.6  &  $+$08 09 01  &  0.028  &  $<$0.102  &     0.155  &   1.339  &     2.212  &  10.72  &  SF  &    5.46  &    8.29  & 0.14 & 0.48 & ~~1/2  \\
  F09395$+$3939  &  09 42 39.2  &  $+$39 26 00  &  0.194  &  $<$0.067  &  $<$0.102  &   0.470  &     1.014  &  12.09  &  Co  &    8.17  &    9.56  & 0.54 & 0.55 & ~~1/2  \\
  F09501$+$5535  &  09 53 31.9  &  $+$55 21 02  &  0.323  &  $<$0.076  &  $<$0.101  &   0.352  &  $<$0.831  &  12.37  &  Co  &    4.17  &    8.32  & 0.77 & 0.71 & ~~1/2  \\
  F10035$+$4852  &  10 06 45.9  &  $+$48 37 44  &  0.065  &     0.098  &     0.283  &   4.593  &     6.242  &  11.93  &  SF  &    6.68  &    7.83  & 0.39 & 0.47 & ~~1/2  \\
  F10300$+$3509  &  10 32 53.9  &  $+$34 53 59  &  0.255  &  $<$0.110  &  $<$0.091  &   0.192  &     0.639  &  12.17  &  Un  &    ...   &    ...   & ...  & ...  & ~~5    \\
  F10565$+$2448  &  10 59 18.1  &  $+$24 32 34  &  0.043  &     0.217  &     1.138  &  12.120  &    15.130  &  11.98  &  Co  &   16.20  &    9.52  & 0.51 & 0.48 & ~~6/7,8 \\
  F11087$+$1036  &  11 11 18.7  &  $+$10 20 20  &  0.047  &  $<$0.188  &  $<$0.164  &   0.745  &     1.248  &  11.00  &  Co  &    5.56  &    8.33  & 0.86 & 0.63 & ~~1/2  \\
  F11163$+$5707  &  11 19 11.2  &  $+$56 50 52  &  0.158  &  $<$0.071  &  $<$0.076  &   0.525  &     0.743  &  11.86  &  SF  &    3.82  &    8.22  & 0.68 & 0.35 & ~~1/2  \\
  F11213$+$6556  &  11 24 24.6  &  $+$65 39 46  &  0.264  &  $<$0.094  &  $<$0.124  &   0.240  &  $<$0.908  &  12.10  &  Co  &    7.04  &    9.07  & 0.90 & 0.80 & ~~1/2  \\
  F11514$+$1212  &  11 54 02.7  &  $+$11 55 28  &  0.365  &  $<$0.123  &  $<$0.131  &   0.236  &  $<$0.967  &  12.45  &  Un  &    ...   &    ...   & ...  & ...  & ~~3    \\
  F12207$+$6329  &  12 23 05.3  &  $+$63 13 21  &  0.059  &  $<$0.125  &     0.308  &   2.160  &     3.310  &  11.57  &  Co  &    4.64  &    8.73  & 1.19 & 0.39 & ~~1/2  \\
  F12232$+$5532  &  12 25 38.3  &  $+$55 15 49  &  0.232  &  $<$0.241  &  $<$0.095  &   0.601  &     0.941  &  12.40  &  Co  &    6.21  &    8.93  & 1.14 & 0.49 & ~~1/2  \\
  F12469$+$4359  &  12 49 16.6  &  $+$43 42 55  &  0.303  &  $<$0.109  &  $<$0.127  &   0.274  &  $<$0.656  &  12.27  &  Co  &    5.62  &    9.17  & 0.53 & 0.51 & ~~1/2  \\
  F12471$+$4759  &  12 49 27.9  &  $+$47 42 50  &  0.304  &  $<$0.118  &  $<$0.067  &   0.357  &  $<$1.168  &  12.36  &  SF  &    4.47  &    8.66  & 0.22 & 0.37 & ~~1/2  \\
  F12479$+$3705  &  12 50 18.2  &  $+$36 49 14  &  0.279  &  $<$0.109  &  $<$0.130  &   0.250  &     0.480  &  12.27  &  Un  &    ...   &    ...   & ...  & ...  & ~~5    \\
  F12536$+$5113  &  12 55 48.4  &  $+$50 57 17  &  0.150  &  $<$0.081  &  $<$0.100  &   0.398  &  $<$1.097  &  11.68  &  SF  &    5.08  &    9.06  & 0.41 & 0.44 & ~~1/2  \\
  F13099$+$4627  &  13 12 06.3  &  $+$46 11 46  &  0.028  &  $<$0.121  &  $<$0.150  &   0.998  &     2.556  &  10.66  &  LI  &    4.44  &    8.25  & 1.72 & 0.83 & ~~1/2  \\
  F13300$+$6652  &  13 31 40.0  &  $+$66 36 34  &  0.104  &  $<$0.088  &  $<$0.075  &   0.424  &  $<$0.795  &  11.33  &  S2  &   15.80  &    9.21  & 2.88 & 1.71 & ~~1/2  \\
  F13380$+$3339  &  13 40 14.4  &  $+$33 24 45  &  0.247  &  $<$0.108  &  $<$0.136  &   0.939  &     1.098  &  12.51  &  S2  &    ...   &    ...   & ...  & ...  & ~~9/10 \\
  F14014$+$3718  &  14 03 37.8  &  $+$37 03 55  &  0.211  &  $<$0.143  &  $<$0.136  &   0.381  &     0.707  &  12.14  &  SF  &    4.83  &    8.73  & 0.85 & 0.37 & ~~1/2  \\
  F14166$+$6514  &  14 17 53.7  &  $+$65 00 25  &  0.364  &  $<$0.069  &  $<$0.056  &   0.189  &  $<$1.488  &  12.35  &  S1  &    ...   &    ...   & ...  & ...  & 11/11 \\
  F14379$+$5420  &  14 39 30.6  &  $+$54 08 07  &  0.268  &  $<$0.064  &  $<$0.053  &   0.333  &     0.428  &  12.20  &  SF  &    5.19  &    8.88  & 0.81 & 0.38 & ~~1/2  \\
  F14413$+$3730  &  14 43 19.6  &  $+$37 18 01  &  0.260  &  $<$0.063  &  $<$0.068  &   0.311  &     0.533  &  12.18  &  Co  &    6.06  &    9.23  & 0.62 & 0.73 & ~~1/2  \\
  F14541$+$4906  &  14 55 49.4  &  $+$48 54 36  &  0.247  &  $<$0.082  &     0.200  &   0.530  &     0.543  &  12.40  &  LI  &    4.05  &    9.15  & 7.51 & 0.56 & ~~1/2  \\
  F15002$+$4945  &  15 01 50.5  &  $+$49 33 38  &  0.337  &  $<$0.053  &     0.074  &   0.395  &     0.510  &  12.53  &  Co  &    5.12  &    9.34  & 0.95 & 0.59 & ~~1/2  \\
  F16533$+$6216  &  16 53 52.1  &  $+$62 11 50  &  0.106  &  $<$0.059  &     0.169  &   1.480  &     2.503  &  11.94  &  Co  &    5.75  &    9.34  & 0.47 & 0.52 & ~~1/2  \\
  F22509$-$0040  &  22 53 33.0  &  $-$00 24 43  &  0.058  &  $<$0.174  &     0.716  &   5.143  &     5.028  &  11.87  &  Co  &    6.34  &    9.07  & 0.66 & 0.63 & ~~1/2  \\
  F23223$+$1459  &  23 24 49.4  &  $+$15 16 32  &  0.014  &     0.144  &  $<$0.339  &   1.310  &     4.461  &  10.29  &  LI  &    3.89  &    7.13  & 1.56 & 1.46 & ~~1/2  \\
   09022$-$3615  &  09 04 12.7  &  $-$36 27 01  &  0.060  &     0.211  &     1.154  &  11.470  &    11.430  &  12.22  &  SF  &    5.67  &    9.27  & 1.16 & 0.33 & 12/4  \\
\enddata
\tablecomments{
Column 1: Object name in the {\it IRAS} catalogs.
Columns 2--3: Right ascension and declination in units of $^{h~m~s}$ and \degr\ \arcmin\ \arcsec, respectively (J2000).
Column 4: Redshift.
Columns 5--8: The {\it IRAS} flux densities at 12, 25, 60, and 100 $\mu$m (Jy).
Column 9: Logarithm of infrared luminosity ($L_\odot$), derived from the  {\it IRAS} fluxes using the formula,
$L_{\rm IR}=2.1\times10^{39}\times D {\rm (Mpc)}^2 \times (13.48\times f_{12}+5.16\times f_{25}+2.58\times f_{60}+f_{100})$, in \citet{san96}.
For sources with upper limits, we follow the method described in \citet{ima08,ima10a}.
The upper and lower limits on the infrared luminosity are obtained
by assuming that the actual flux is equal to the {\it IRAS} upper limit and zero value, respectively,
and the average of these values is adopted as the infrared luminosity.
Column 10: Optical spectral type, determined with spectroscopic data covering at least
from H$\beta$+[O{\sc iii}]$\lambda5007$ to H$\alpha$+[N{\sc ii}]$\lambda6584$
using the criteria of \citet{kew06}
(SF=star-forming; Co=composite; LI=LINER; S2=Seyfert 2; S1=Seyfert 1; Un=Unclassified).
Column 11: Observed H$\alpha$/H$\beta$ ratio.
Column 12: Logarithm of H$\alpha$ luminosity ($L_\odot$), corrected for dust extinction using the observed H$\alpha$/H$\beta$ ratio
with the \citet{cal00} extinction law ($R_V=4.05$) on assuming the intrinsic H$\alpha$/H$\beta$ ratio of 2.85 for SF galaxies
and 3.1 for composite plus AGN galaxies \citep{ost06}.
For SDSS (U)LIRGs in \citet{hwa10a},
the aperture correction is also applied following the method suggested by \citet{hop03}
based on the difference between fiber and Petrosian magnitudes to minimize the small fixed-size (3\arcsec) aperture effect, if available.
Column 13: [O{\sc iii}]$\lambda5007$/H$\beta$ ratio, corrected for dust extinction.
Column 14: [N{\sc ii}]$\lambda6584$/H$\alpha$ ratio, corrected for dust extinction.
Column 15: References of the identification/optical properties.
(1) \citealt{hwa10a}; (2) This study; (3) \citealt{hwa07}; (4) \citealt{lee11}; (5) \citealt{sta00}; (6) \citealt{soi89};
(7) \citealt{vei95}; (8) \citealt{yua10}; (9) \citealt{lee94}; (10) \citealt{ver06}; (11) \citealt{hou09}; (12) \citealt{san03}.
}
\end{deluxetable}

\clearpage
\begin{deluxetable}{rlll}
\tabletypesize{\scriptsize}
\tablecaption{Observation log.\label{table2}}
\tablewidth{0pt}
\tablehead{
\colhead{Object} &
\colhead{Program} &
\colhead{Observation ID} &
\colhead{Observation Date}
}
\startdata
F01159$-$0029  & CLNSL &  3350035-001--002\tablenotemark{a}  & 2009-07-09 \\
F01212$-$5025  & NULIZ &  3050012-001--002                   & 2006-12-12 \\
F02595$-$4714  & NULIZ &  3050008-001--002                   & 2007-01-07--08 \\
F03130$-$3119  & NULIZ &  3050009-001--002                   & 2007-01-25 \\
F03483$-$4704  & NULIZ &  3050010-001--002                   & 2007-01-22 \\
F08082$+$2521  & NISIG &  3610001-001                        & 2009-10-22 \\
F09019$+$5136  & NISIG &  3610011-001--002                   & 2009-10-25--26 \\
F09174$+$0821  & NISIG &  3610035-001--002                   & 2009-11-12--13 \\
F09395$+$3939  & CLNSL &  3350003-001--003                   & 2009-05-05 \\
F09501$+$5535  & NULIZ &  3050001-001--002                   & 2006-11-01, 2007-04-29 \\
F10035$+$4852  & NISIG &  3610008-001--002                   & 2009-11-06--07 \\
F10300$+$3509  & NULIZ &  3051003-001--002                   & 2007-05-17 \\
F10565$+$2448  & NULIZ &  3051019-001                        & 2007-05-28 \\
F11087$+$1036  & CLNSL &  3350028-001                        & 2009-06-05 \\
F11163$+$5707  & NISIG &  3610013-001--004                   & 2009-11-14 \\
F11213$+$6556  & NULIZ &  3050030-001                        & 2007-05-04 \\
F11514$+$1212  & NULIZ &  3051008-001--002                   & 2007-06-15 \\
F12207$+$6329  & CLNSL &  3350024-001--002                   & 2009-05-14 \\
F12232$+$5532  & NULIZ &  3051006-001--002                   & 2007-05-25 \\
F12469$+$4359  & NULIZ &  3051010-001--002                   & 2007-06-11 \\
F12471$+$4759  & NULIZ &  3051009-001--002                   & 2007-06-07--08 \\
F12479$+$3705  & NULIZ &  3051005-001--002                   & 2007-06-16 \\
F12536$+$5113  & NISIG &  3610028-001--004                   & 2009-12-06--07 \\
F13099$+$4627  & CLNSL &  3350015-001                        & 2009-06-12 \\
F13300$+$6652  & CLNSL &  3350025-001--008\tablenotemark{b}  & 2008-11-19--22, 2009-05-18--20 \\
F13380$+$3339  & NULIZ &  3051001-001--002                   & 2007-06-30 \\
F14014$+$3718  & NULIZ &  3051007-001--002                   & 2007-07-03 \\
F14166$+$6514  & NULIZ &  3051014-001--002                   & 2007-05-28 \\
F14379$+$5420  & NULIZ &  3050004-001--002                   & 2006-12-24 \\
F14413$+$3730  & NULIZ &  3050005-001--002                   & 2007-01-11--12 \\
F14541$+$4906  & NULIZ &  3050006-001--002                   & 2007-01-04 \\
F15002$+$4945  & NULIZ &  3051015-001--002                   & 2007-07-06 \\
F16533$+$6216  & CLNSL &  3350023-001--002                   & 2009-01-08 \\
F22509$-$0040  & CLNSL &  3350036-001--002                   & 2009-06-04 \\
F23223$+$1459  & CLNSL &  3350037-001--005                   & 2009-06-19--20 \\
 09022$-$3615  & NULIZ &  3051018-001--002                   & 2007-05-26
\enddata
\tablenotetext{a}{The 2nd pointing was not used to obtain the final spectra.}
\tablenotetext{b}{The 4th pointing was not used to obtain the final spectra.}
\end{deluxetable}

\clearpage
\begin{deluxetable}{rrrrrrrrrc}
\tabletypesize{\scriptsize}
\tablecaption{Measurements and AGN signature.\label{table3}}
\tablewidth{0pt}
\tablehead{
\colhead{Object} &
\colhead{Aper.} &
\colhead{$f_{\rm 3.3 PAH}$} &
\colhead{EW$_{\rm 3.3 PAH}$} &
\colhead{$f_{\rm Br\alpha}$} &
\colhead{$f_{\rm Br\beta}$} &
\colhead{$\tau _{3.1}$} &
\colhead{$\tau _{3.4}$} &
\colhead{$\Gamma$} &
\colhead{AGN} \\
\colhead{(1)} & \colhead{(2)} & \colhead{(3)} & \colhead{(4)} & \colhead{(5)} & \colhead{(6)} & \colhead{(7)} & \colhead{(8)} & \colhead{(9)} & \colhead{(10)}
}
\startdata
  F01159$-$0029   & 8.8 &       6.9  $\pm$  1.8 &       59  $\pm$  16 &              $<$  2.0 &       2.6  $\pm$  1.1 &              $<$  0.10 &              ...       &  $-$2.72  $\pm$ 0.07 &  X           \\
  F01212$-$5025   &10.2 &              $<$  0.5 &             $<$  13 &              ...      &              $<$  0.5 &              $<$  0.15 &              $<$  0.14 &  $-$2.06  $\pm$ 0.06 &  $\bigcirc$  \\
  F02595$-$4714   & 8.8 &       0.3  $\pm$  0.1 &       18  $\pm$  ~6 &              ...      &       0.8  $\pm$  0.1 &              $<$  0.25 &              $<$  0.35 &  $-$2.13  $\pm$ 0.07 &  $\bigcirc$  \\
  F03130$-$3119   & 8.8 &       1.2  $\pm$  0.1 &       81  $\pm$  ~9 &              ...      &              $<$  0.6 &              $<$  0.39 &              $<$  0.15 &  $-$2.38  $\pm$ 0.10 &  X           \\
  F03483$-$4704   & 8.8 &              $<$  0.5 &             $<$  28 &              ...      &              $<$  0.6 &              $<$  0.43 &              $<$  0.24 &  $-$1.58  $\pm$ 0.10 &  $\bigcirc$  \\
  F08082$+$2521   &13.1 &       6.6  $\pm$  1.7 &       40  $\pm$  10 &              ...      &              ...      &              $<$  0.13 &              ...       &  $-$3.11  $\pm$ 0.08 &  X           \\
  F09019$+$5136   &10.2 &      15.4  $\pm$  2.4 &      104  $\pm$  16 &              $<$  1.8 &              $<$  1.4 &              $<$  0.12 &              ...       &  $-$2.37  $\pm$ 0.06 &  X           \\
  F09174$+$0821   &11.7 &      10.1  $\pm$  1.6 &       48  $\pm$  ~7 &              $<$  1.0 &              $<$  2.1 &              $<$  0.10 &              ...       &  $-$2.66  $\pm$ 0.05 &  X           \\
  F09395$+$3939   & 7.3 &       1.2  $\pm$  0.2 &       60  $\pm$  ~8 &              $<$  0.7 &              $<$  0.8 &              $<$  0.13 &              ...       &  $-$2.48  $\pm$ 0.14 &  X           \\
  F09501$+$5535   & 7.3 &       0.5  $\pm$  0.1 &       58  $\pm$  ~8 &              ...      &              $<$  0.4 &       0.96 $\pm$  0.16 &      0.59  $\pm$  0.30 &  $-$1.00  $\pm$ 0.09 &  $\bigcirc$  \\
  F10035$+$4852   & 8.8 &      14.6  $\pm$  1.3 &       83  $\pm$  ~8 &              $<$  1.3 &              $<$  2.0 &              $<$  0.13 &              ...       &  $-$2.53  $\pm$ 0.05 &  X           \\
  F10300$+$3509   & 7.3 &       0.5  $\pm$  0.1 &       43  $\pm$  11 &              ...      &              $<$  0.6 &              $<$  0.38 &              ...       &  $-$2.61  $\pm$ 0.15 &  X           \\
  F10565$+$2448   &10.2 &      37.0  $\pm$  5.6 &      111  $\pm$  17 &       4.4  $\pm$  0.8 &       4.9  $\pm$  1.4 &       0.36 $\pm$  0.06 &              ...       &  $-$2.30  $\pm$ 0.05 &  $\bigcirc$  \\
  F11087$+$1036   &10.2 &       5.6  $\pm$  1.1 &       88  $\pm$  18 &              $<$  0.8 &              $<$  1.6 &              $<$  0.14 &              $<$  0.14 &  $-$3.07  $\pm$ 0.11 &  X           \\
  F11163$+$5707   & 7.3 &       1.5  $\pm$  0.3 &       66  $\pm$  14 &              $<$  0.9 &              $<$  1.0 &              $<$  0.15 &              ...       &  $-$1.97  $\pm$ 0.15 &  X           \\
  F11213$+$6556   & 7.3 &              $<$  0.5 &             $<$  37 &              ...      &              $<$  0.8 &              $<$  0.26 &              $<$  0.27 &  $-$2.26  $\pm$ 0.16 &  $\bigcirc$  \\
  F11514$+$1212   & 5.8 &              $<$  0.5 &             $<$  61 &              ...      &              $<$  0.5 &              $<$  0.15 &              ...       &  $-$4.54  $\pm$ 0.22 &  X           \\
  F12207$+$6329   &10.2 &      15.7  $\pm$  1.8 &      180  $\pm$  20 &       3.0  $\pm$  0.3 &       2.3  $\pm$  0.4 &              $<$  0.09 &              ...       &  $-$2.79  $\pm$ 0.08 &  X           \\
  F12232$+$5532   & 8.8 &       1.0  $\pm$  0.1 &       39  $\pm$  ~3 &              ...      &              $<$  0.7 &              $<$  0.16 &              $<$  0.21 &  $-$1.86  $\pm$ 0.08 &  $\bigcirc$  \\
  F12469$+$4359   & 7.3 &              $<$  0.6 &             $<$  88 &              ...      &              $<$  0.5 &       1.21 $\pm$  0.47 &              ...       &  $-$3.45  $\pm$ 0.19 &  $\bigcirc$  \\
  F12471$+$4759   & 7.3 &       0.6  $\pm$  0.2 &       22  $\pm$  ~7 &              ...      &              $<$  0.6 &              $<$  0.20 &              $<$  0.14 &  $-$2.01  $\pm$ 0.07 &  $\bigcirc$  \\
  F12479$+$3705   & 7.3 &       0.6  $\pm$  0.2 &       28  $\pm$  ~8 &              ...      &              $<$  0.6 &       0.22 $\pm$  0.07 &              $<$  0.11 &  $-$2.53  $\pm$ 0.10 &  $\bigcirc$  \\
  F12536$+$5113   & 7.3 &       4.0  $\pm$  0.6 &      147  $\pm$  23 &              $<$  1.6 &              $<$  1.2 &              $<$  0.25 &              ...       &  $-$2.62  $\pm$ 0.28 &  X           \\
  F13099$+$4627   &11.7 &              $<$  5.1 &             $<$  15 &              $<$  1.2 &              $<$  3.2 &       0.12 $\pm$  0.04 &              ...       &  $-$3.20  $\pm$ 0.05 &  $\bigcirc$  \\
  F13300$+$6652   & 7.3 &              $<$  0.3 &             $<$  22 &              $<$  0.3 &              $<$  0.5 &              $<$  0.24 &              $<$  0.23 &  $-$2.93  $\pm$ 0.11 &  $\bigcirc$  \\
  F13380$+$3339   & 7.3 &       1.1  $\pm$  0.1 &       44  $\pm$  ~5 &              ...      &              $<$  0.6 &              $<$  0.35 &              $<$  0.15 &  $-$2.16  $\pm$ 0.08 &  X           \\
  F14014$+$3718   & 7.3 &       1.4  $\pm$  0.1 &       76  $\pm$  ~5 &              ...      &              $<$  0.7 &              $<$  0.14 &              ...       &  $-$1.68  $\pm$ 0.08 &  X           \\
  F14166$+$6514   & 5.8 &              $<$  0.7 &             $<$  31 &              ...      &              $<$  0.5 &              $<$  0.18 &              ...       &  $-$2.04  $\pm$ 0.20 &  $\bigcirc$  \\
  F14379$+$5420   & 4.4 &       0.6  $\pm$  0.2 &       71  $\pm$  27 &              ...      &              $<$  0.3 &              $<$  0.42 &              ...       &  $-$1.86  $\pm$ 0.21 &  X           \\
  F14413$+$3730   & 7.3 &       0.9  $\pm$  0.3 &       55  $\pm$  20 &              ...      &              $<$  0.5 &              $<$  0.37 &              $<$  0.12 &  $-$2.54  $\pm$ 0.14 &  X           \\
  F14541$+$4906   & 8.8 &              $<$  0.4 &             $<$  ~3 &              ...      &              $<$  0.7 &              $<$  0.08 &       0.29  $\pm$ 0.05 &     0.62  $\pm$ 0.07 &  $\bigcirc$  \\
  F15002$+$4945   & 7.3 &              $<$  0.7 &             $<$  ~3 &              ...      &              $<$  0.7 &              $<$  0.11 &              ...       &  $-$0.52  $\pm$ 0.05 &  $\bigcirc$  \\
  F16533$+$6216   &10.2 &       6.7  $\pm$  0.8 &      137  $\pm$  16 &       0.6  $\pm$  0.3 &       0.4  $\pm$  0.2 &       0.42 $\pm$  0.14 &              ...       &  $-$2.33  $\pm$ 0.08 &  $\bigcirc$  \\
  F22509$-$0040   & 7.3 &      17.8  $\pm$  1.9 &      125  $\pm$  15 &       3.0  $\pm$  1.2 &              $<$  2.1 &       0.62 $\pm$  0.19 &       0.38  $\pm$ 0.14 &  $-$0.91  $\pm$ 0.08 &  $\bigcirc$  \\
  F23223$+$1459   &13.1 &              $<$  1.5 &             $<$  ~3 &              $<$  0.9 &              ...      &       0.23 $\pm$  0.06 &              ...       &  $-$3.38  $\pm$ 0.05 &  $\bigcirc$  \\
   09022$-$3615   &13.1 &      21.0  $\pm$  4.4 &       44  $\pm$  10 &       4.9  $\pm$  0.9 &       4.3  $\pm$  1.2 &              $<$  0.08 &              $<$  0.06 &     0.55  $\pm$ 0.12 &  $\bigcirc$  \\
\enddata
\tablecomments{
Column 1: Object name in the {\it IRAS} catalogs.
Column 2: Aperture width used for the extraction (\arcsec).
Column 3: Observed flux of the 3.3 $\mu$m PAH emission ($10^{-14}$ ergs s$^{-1}$ cm$^{-2}$).
Column 4: Rest-frame equivalent width of the 3.3 $\mu$m PAH emission (nm).
Columns 5--6: Observed fluxes of Br$\alpha$ and Br$\beta$ ($10^{-14}$ ergs s$^{-1}$ cm$^{-2}$).
The Br$\alpha$ and Br$\beta$ fluxes of sources at z $>$ 0.2 and z $<$ 0.02 are not presented, respectively,
because the measurements near the edge of the spectra are uncertain.
Columns 7--8: Optical depths of the 3.1 $\mu$m H$_{2}$O ice and 3.4 $\mu$m bare carbonaceous dust absorption features.
The 3.4 $\mu$m optical depths are not presented if they are significantly affected by the 3.4 $\mu$m PAH sub-peak.
Column 9: Continuum slope $\Gamma$ ($F_\lambda$ $\varpropto$ $\lambda^{\Gamma}$).
Column 10: AGN signature from the NIR features
(EW$_{\rm 3.3 PAH}<$ 40 nm, $\tau _{3.1}>$ 0.3, $\tau _{3.4}>$ 0.2, or $\Gamma> -$1).
}
\end{deluxetable}

\clearpage
\begin{deluxetable}{ccr@{~}lr@{~}lr@{~}lr@{~}lr@{~}lr@{~}l}
\tabletypesize{\scriptsize}
\tablecaption{NIR AGN detection rate in bins of optical spectral type and of infrared luminosity.\label{table4}}
\tablewidth{0pt}
\tablehead{
\colhead{Sample} &
\colhead{log $L_{\rm IR}$} &
\multicolumn{12}{c}{Optical spectral type} \\
\cline{3-14}
 & ($L_{\odot}$) & \multicolumn{2}{c}{Star-forming} &  \multicolumn{2}{c}{Composite} &  \multicolumn{2}{c}{LINER} &
                   \multicolumn{2}{c}{Seyfert 2}    &  \multicolumn{2}{c}{Seyfert 1} &  \multicolumn{2}{c}{Unclassified}}
\startdata
This study     & 12.3--13.0 & 100\% & (1/1)  & 100\% & (3/3)   & ...   & (0/0) & 50\%  & (1/2)   & 100\% & (1/1) & 0\%  & (0/1) \\
               & 12.0--12.3 & 50\%  & (2/4)  & 60\%  & (3/5)   & ...   & (0/0) & ...   & (0/0)   & ...   & (0/0) & 50\% & (2/4) \\
               & 11.0--12.0 & 0\%   & (0/3)  & 50\%  & (3/6)   & ...   & (0/0) & 100\% & (1/1)   & ...   & (0/0) & ...  & (0/0) \\
               & 10.0--11.0 & 0\%   & (0/3)  & ...   & (0/0)   & 100\% & (2/2) & ...   & (0/0)   & ...   & (0/0) & ...  & (0/0) \\
\cline{1-14}\\[-8pt]
Imanishi+08,10 & 12.3--13.0 & 100\% & (2/2)  & 77\%  & (10/13) & ...   & (0/0) & 75\%  & (6/8)   & ...   & (0/0) & 0\%  & (0/2) \\
               & 12.0--12.3 & 67\%  & (4/6)  & 61\%  & (19/31) & 50\%  & (1/2) & 70\%  & (14/20) & ...   & (0/0) & 57\% & (4/7) \\
               & 11.0--12.0 & 18\%  & (3/17) & 25\%  & (5/20)  & ...   & (0/0) & 50\%  & (5/10)  & 100\% & (1/1) & 20\% & (2/10)
\enddata
\tablecomments{The values in parentheses mean the number of sources with AGN signature/total number of sources in each bin
from our sample and the sample in \citet{ima08,ima10a}.
The optical spectral types of the Imanishi sample are taken from \citet{yua10}
so that the same criteria of \citet{kew06} are used for the two samples.
For sources with multiple nuclei, if AGN signatures are seen at least in one nucleus, then these sources are considered to be AGN.
For an unbiased comparison, the additional interesting sources in the Imanishi sample are not included.
}
\end{deluxetable}

\end{document}